%% file: main.tex
\documentclass[a4paper, 10pt]{extarticle}
\usepackage[utf8]{inputenc}
\usepackage{natbib}
\usepackage{graphicx}
\usepackage{physics}
\usepackage{amsmath}
\usepackage{amsfonts}
\usepackage[margin=1in]{geometry}
\usepackage{float}
\usepackage{pgfplots}
\usepackage{multicol}
\usepackage{geometry}
\usepackage{hyperref}
\usepackage{verbatim}
\usepackage{listings}
\usepackage{tcolorbox}
\usepackage{lipsum}
\usepackage{courier}
\usepackage{bbold}
\usepackage{natbib}
\usepackage{epigraph}
\usepackage{authblk}

\numberwithin{equation}{subsection}

\title{Fractional Revival Dynamics in Kerr-Type Systems: Angular Momentum Moments and Classical Analogs}

\begin{document}

\author[1]{Ashish Kumar Patra%
\thanks{Present address: QClairvoyance Quantum Labs Pvt. Ltd., Secunderabad, TG 500094, India.
}}

\author[2]{Saikumar Krithivasan%
\thanks{Corresponding author. Email: saikumarphy93@gmail.com}}

\affil[1]{Department of Physics, Loyola College, TN 600034, India}
\affil[2]{Department of Physics, D. G. Vaishnav College, TN 600106, India}

\date{}
\maketitle

\section*{Abstract}

Wave packet revivals and fractional revivals are hallmark quantum interference phenomena that arise in systems with nonlinear energy spectra, and their signatures in expectation values of observables have been studied extensively in earlier work. In this article, we build on these studies and extend the analysis in two important directions. First, we investigate fractional revival dynamics in angular momentum observables, deriving explicit expressions for the time evolution of their moments and demonstrating that higher-order angular momentum moments provide clear and selective signatures of fractional revivals. Second, we examine classical analogs of quantum revival phenomena and elucidate structural similarities between quantum fractional revivals and recurrence behavior in representative classical systems. Using the Kerr-type nonlinear Hamiltonian as a paradigmatic model, we analyze the autocorrelation function, moment dynamics, and phase-space structures, supported by visualizations such as quantum carpets. Our results broaden the range of experimentally accessible diagnostics of fractional revivals and provide a unified perspective on revival phenomena across quantum and classical dynamical systems.

% \tableofcontents

\section{INTRODUCTION}

% \epigraph{\textbf{\textit{"...I can safely say that nobody understands Quantum Mechanics."}}}{Dr. Richard Feynman}

Since the conception of the field of Quantum Mechanics, physicists and scientists have been baffled by the implications of the subject. Many of the seemingly anomalous effects like quantum superposition~\cite{griffiths_introduction_quantum_mechanics_2ed}, collapse of a quantum state upon observation, wave-particle duality are still held to be postulates in the subject, while they are being debated by physicists.One such interesting property is the concept of \textbf{Revivals and Fractional Revivals}\cite{sudheesh_2004_frac_revivals} \cite{Robinett_2004}. This phenomenon has been studied by scientists owing to the vast amount of applications one can use it for. Simply put, when a quantum system is left unhindered (\textit{a closed quantum system} , so to speak) and allowed to evolve in time, the system loses its initial properties but at a later time, comes back to its original state. This phenomenon is dubbed as the \textit{"revival"} of the quantum system, and the time required for it to come back to its original state is called the \textit{"revival time"}.
\par In this work, we revisit the phenomenon of quantum revivals and fractional revivals from the perspective of
\textit{observable-based diagnostics}, with particular emphasis on \textbf{angular momentum observables and
their higher-order moments}. While earlier studies~\cite{sudheesh_2004_frac_revivals}~\cite{Rohith_2015}~\cite{M_Nauenberg_1990} have primarily focused on wave-function overlap measures
such as the autocorrelation function, or on expectation values of position- and momentum-like observables,
a systematic analysis of revival signatures encoded in angular momentum moments has received comparatively
less attention. Motivated by both conceptual clarity and experimental accessibility, we demonstrate that
higher moments of angular momentum operators provide clear, selective, and robust signatures of fractional
revival dynamics, often revealing structures that remain hidden at the level of lower-order observables.

In addition to the quantum analysis, we also explore \textit{classical analogs} of revival and fractional revival
phenomena. By examining representative classical systems exhibiting recurrence and aliasing-like behavior,
we highlight structural similarities between quantum fractional revivals and classical recurrence patterns.
This parallel viewpoint helps place revival phenomena within a broader dynamical context and clarifies which
aspects are intrinsically quantum and which admit classical interpretation.

The paper is organized as follows.
In \textbf{Section~2}, we introduce the theoretical framework for revivals and fractional revivals, focusing on
wave-packet dynamics in nonlinear systems and using the Kerr-type Hamiltonian as a paradigmatic model. We
review the role of spectral nonlinearity, discuss revival time scales, and analyze standard diagnostics such as
the autocorrelation function.
In \textbf{Section~3}, we present the central results of this work: we derive explicit expressions for the time
evolution of angular momentum observables and their moments, and demonstrate how higher-order moments act
as sensitive indicators of fractional revivals. Numerical simulations and visualizations, including phase-space
structures and quantum carpets, are used to illustrate these signatures.
In \textbf{Section~4}, we examine classical analogs of revival phenomena, drawing connections between quantum
fractional revivals and recurrence behavior in classical dynamical systems.
Finally, \textbf{Section~5} summarizes our results and discusses possible extensions and experimental
implications of the observable-based approach developed here.

\section{Theoretical Background}

\subsection{Relation between the phase velocity and group velocity of a wavepacket}

\par We know that a quantum particle can be approximated to look like a "wavepacket". A wavepacket is defined as an infinite superposition of sinusoidal waves, with varying wavenumbers and frequencies, which move as a single unit in time. These individual sinusoidal waves can be linked to individual wavefunctions corresponding to allowed energy levels (in which case, the frequencies and wavenumbers are quantized, and increase in an integral linear or non-linear fashion). The combined propagation of these individual waves with individual velocities (which we shall call phase velocity), makes the whole gaussian wave packet propagate forward with a velocity called the group velocity, which is the velocity of the quantum particle.

   We shall briefly touch upon the concepts of phase velocity, group velocity and dispersion relation to get an idea about the dynamics of a wave-packet.
   Let us consider an individual sine wave from the infinite superposition of the waves which make a wave-packet, whose wave-number (or number of waves in unit length) is $k$, and frequency of oscillation is $\omega$:
   \begin{equation}
       sin(kx - \omega t)
   \end{equation}
   We know that wave-number(one can also call the wave-number as the spatial frequency of the wave) is inversely proportional to the wavelength $\lambda$, and frequency inversely proportional to the time period of oscillation $\tau$. The speed of the wave is determined by its phase $``\theta" = kx - \omega t$.
   
Then, the phase velocity, or the velocity of an individual wave can be stated as the following:

\begin{equation}
    v_p = \frac{\lambda}{\tau} = \frac{\omega}{k}
\end{equation}

The velocity of the wave-packet as a whole depends on the phase velocity of each individual wave.The velocity of the wave-packet, the group velocity $v_g$, is determined by the variation of the frequency of the waves with respect to the variation in the wave-number, as we assumed that the wave-packet is made up of infinite sinusoidal waves.

\begin{equation}
    v_g = \frac{\partial{\omega(k)}}{\partial{k}}
\end{equation}
Here $\omega(k)$ is called the \textit{"dispersion relation"}. $\omega(k)$ corresponds to the distribution of the individual wave-numbers and in our case is assumed to be a continuous function. We see that if $\omega(k) \propto k$, then $v_g$ will be a constant value. If the phase velocity of the constituent waves are constant with respect to time, then $\dot{v_p} = 0$:
\begin{equation}
v_p = \frac{d{\theta}}{dt} = \frac{d(kx - \omega t)}{dt}= k \frac{dx}{dt} - \omega = 0
\end{equation}

\begin{equation}
    \frac{dx}{dt} = \frac{\omega}{k} = constant
\end{equation}

This implies that if the dispersion relation is directly proportional to $k$, then the wave-packet propagates forward without any dispersion, and its group velocity is equal to its phase velocity. Its overall shape might change, but the wave-packet remains intact.
A photon wave-packet propagating through vacuum, for example, has a dispersion relation directly proportional to the wave-number, and thus the wave-packet does not
spread as it moves forward.

\par
If the dispersion relation is directly proportional to higher powers of $k$, i.e., $\omega(k) \propto k^n$, where $n>1$, then we see that the group velocity no longer remains constant, but becomes a function of the wave-number itself. In other words, the wave-packet spreads and gets distorted as it moves forward in space and time.

\subsection{Time Evolution of Quantum States}

\par If we observe a system to be in a normalized state $\ket{\psi}$ at some time $t_0$ there is no necessity for it to remain the same after some time interval. To obtain the state $\ket{\psi}$ at some later time, say, $t^\prime$, we will introduce an operator $\hat{S}(t)$(which we shall as of yet call the "Time Evolution Operator"), which will give us the state at some time $t$, that is:

\begin{equation}
    \hat{S} \ket{\psi(x, t = t_0)} = \ket{\psi(x,t = t^\prime)}
\end{equation}
As the system only evolves in time, and remains otherwise unaffected(a closed quantum system), the total probability of finding the particle in the system must be equal to one.Thus, it is only reasonable for us to set the Time Evolution Operator to be a unitary operator.
\begin{equation}
    \hat{S} \hat{S^{\dagger}} = 1
\end{equation}
The Time Dependent Schrodinger's equation is given as:
\begin{equation}
    i\hbar \frac{\partial{\psi(\vec{r}, t)}}{\partial t} = -\frac{\hbar^2}{2m} \frac{\partial^2{\psi(\vec{r}, t)}}{\partial^2 x} + V\psi(\vec{r}, t)
\end{equation}

When the quantum system has no form of interactions with the environment, the absolute square of the sum of these energy eigen values remain constant(if normalized, unity), as the energy operator $\hat{H}$ is not a function of time ($\hat{H} \neq \hat{H}(t)$). As the energy operator which is an eigen operator of the wave function remains constant, one can rewrite the above equation as the following:

\begin{equation}
   i\hbar \frac{\partial{\psi(\vec{r}, t)}}{\partial{t}} = \hat{H} \psi 
\end{equation}

The above equation might have been a straight-forward first order differential equation, if not for the imaginary term in the equation. Without it, the equation would have yielded a solution pertaining to decay. The solution for an equation of form

\begin{equation}
\dfrac{\partial{y(x)}}{\partial{x}} = c\cdot y(x) 
\end{equation}

would look like $ y(x) = Ce^x + D e^{-x} $. 
But the differential equation: 

\begin{equation}
i \dfrac{\partial f(x)}{\partial x} = c\cdot f(x)
\end{equation}
has a solution of form $y(x) = A e^{ix} + B e^{-ix}$, which pertains to a sinusoidal variation.Hence the system always remain within a finite range.

One way to look at it intuitively is that the function y(x) in equation (1.2.5) is a diffusion equation, where in, the rate of change of the value is dependent on the initial condition in a direct manner. That is, for large values of 'x'(let's say, x = $x'$, where, $x' \longrightarrow \infty$), the function y($x'$) will of course, take large values, \textbf{and}, the corresponding rate of change of y(x) in the vicinity of $x'$ also \textit{will be large.}

Looking at equation (1.2.6) in the similar vein, we would initially say that it also must be a "sort of" diffusion equation. But its solution, as we see, tells us that for any possible value of x, the \textbf{rate of change} of f(x) is \textit{does not increase or decrease according to the magnitude of x}. The presence of a complex term in the equation is the cause of this sort of behavior.Then where does the diffusion take place? \textbf{In the complex plane!!!} So one way to understand equation (1.2.6) is that it is a diffusion equation, and the diffusion takes place in the imaginary plane.

The solution of the equation no 1.2.4 is thus: (here, we have assumed that the function $\psi$ is the product of two separate functions pertaining to position and time.)

\begin{equation}
    \psi (\vec{r}, t) = \phi(\vec{r}) e^{-i\frac {\hat{H}t}{\hbar}}
\end{equation}

The operator $\hat{H}$ determines the way in which the system evolves in time

\begin{equation}
    \hat{S} = e^{-i \frac{\hat{H}t}{\hbar}}
\end{equation}

And the above operator when operated onto a state will tell us how the state will "look"(depending on the representation we are using to describe the wave function, usually it is in Fock space) like after some time t.

The other intriguing way in which we can look at time evolution of states is in the following manner.
   We know that in a mathematical sense, states are nothing but infinite dimensional vectors(either row matrices or column matrices, depending), and their corresponding operators are square matrices.
   
Thus the time evolution operator is the exponential of a complex term along with a matrix, i.e., something of form:

\begin{equation}
    e^{-i \tiny{\begin{bmatrix}a & b &..\\ c & d &.. \\ ..&..&..\end{bmatrix}}t}
\end{equation}

Which is nothing but :
\begin{equation}
        e^{-i \tiny{\begin{bmatrix}a & b &..\\ c & d &.. \\ ..&..&..\end{bmatrix}}t} = \mathbb{1} + it\begin{bmatrix}a & b &..\\ c & d &.. \\ ..&..&..\end{bmatrix}- \frac{t^2}{2!} \begin{bmatrix}a & b &..\\ c & d &.. \\ ..&..&..\end{bmatrix}^2 - \frac{i t^3}{3!} \begin{bmatrix}a & b &..\\ c & d &.. \\ ..&..&..\end{bmatrix}^3 +...
\end{equation}

Which upon calculation will converge into a specific matrix, which will be a function of time t. We can be sure of it's convergence due to the complex term along with the matrix. Thus, we see that the dimensions of the square matrix, remain unchanged in this operation, and when the resultant matrix is multiplied with a state vector (a row or column matrix), it will correspond to a \textbf{unitary transformation} in the infinite dimensional space.

\subsection{Ladder Operators and Number Operators- An Overview}
\par
The introduction of ladder operators led to an elementary and intuitive understanding and solving of the Quantum Harmonic Oscillator, as opposed to the analytical method, which employs rigorous mathematical tools to obtain the same result.

The equation to be solved in the case of QHO, is:
\begin{equation}
    \frac{1}{2m}[p^2 + (m\omega x)^2]\psi = E\psi
\end{equation}
To solve it algebraically, these two dimensionless quantities were introduced:
\begin{equation}
    a^{\dagger} = \frac{1}{\sqrt{2\hbar m\omega}} (-ip + m\omega x)
\end{equation}
\begin{equation}
    a = \frac{1}{\sqrt{2\hbar m \omega}} (ip + m\omega x)
\end{equation}
From the above two equations, we can obtain the values of x and p, as the following:
\begin{equation}
    x = \sqrt{\frac{\hbar}{2m\omega}}(a + a^{\dagger})
\end{equation}
\begin{equation}
    p = \frac{1}{i} \sqrt{\frac{\hbar m \omega}{2}}(a - a^{\dagger})
\end{equation}
We can readily obtain the commutation relation between $a$ and $a^{\dagger}$ is:
\begin{equation}
    [a,a^{\dagger}] = aa^{\dagger} - a^{\dagger}a = 1
\end{equation}

The product of the operators $a$ and $a^{\dagger}$ , is given as:
\begin{equation}
    aa^{\dagger} = \frac{1}{2\hbar m \omega} [p^2 + {(m\omega x)}^2] - \frac{i}{2\hbar} [x,p]
\end{equation}
Using the commutation relation $[x,p] = i\hbar$, we get the Hamiltonian as:
\begin{equation}
    H = \hbar \omega(a^\dagger a + \frac{1}{2})
\end{equation}

The operators $a^\dagger$ and $a$ are asserted to perform the following operations when acted upon on a stationary state.
\begin{equation}
    a \ket{n} = \sqrt{n} \ket{n-1}
\end{equation}
and:
\begin{equation}
    a^\dagger \ket{n} = \sqrt{n+1} \ket{n+1}
\end{equation}

Repeated operations of the raising operator on the ground state $\ket{0}$ and the lowering operator on some arbitrary state $\ket{n}$  can be written as the following:
\begin{equation}
    \ket{n} = \frac{1}{\sqrt{n!}} a^{\dagger n} \ket{0} \quad ; \quad \ket{0} = \frac{1}{\sqrt{n!}} a^n \ket{n}
\end{equation}

Here we would like to consider the combination $a^\dagger a$ which when operated on a stationary state, say $n$ gives us:
\begin{equation}
    a^\dagger a \ket{n} =  (\sqrt{n}) a^\dagger \ket{n - 1} = n \ket{n}
\end{equation}

Hence we see that the combined operation gives the "number" of the energy state. So, it is appropriate to call this the Number Operator $\hat{N}$.

$\hat{N}$ does not commute with $a$ and $a^\dagger$:

\begin{equation}
[\hat{N}, a] = [a^\dagger a, a] = a^\dagger[a,a] + [a^\dagger,a]a=-a    
\end{equation}

\begin{equation}
    [\hat{N}, a^\dagger] = [a^\dagger a,a^\dagger] = a^\dagger [a, a^\dagger] +[a^\dagger,a^\dagger] a = a^\dagger
\end{equation}

If a Hamiltonian $\hat{H}$, commutes with the number operator, then the Hamiltonian can be written as a function of the number operator. Consider the Hamiltonian $\hat{H} = {\hbar \omega} (a^{\dagger}a + \frac{1}{2})$

\subsection{Coherent States}
\subsubsection{Introduction}
There are certain linear combinations of the energy eigen-states of the quantum harmonic oscillator, which are eigen-values of the lowering operator $a$.\newline These linear combinations of the stationary states are collectively called as a coherent state, and are most commonly represented as $\ket{\alpha}$. \par
The eigenvalues of these coherent states are represented by $\alpha$.

And so, simply put:

\begin{equation}
    a \ket{\alpha} = \alpha \ket{\alpha}
\end{equation}

Similarly,
\begin{equation}
    \bra{\alpha}\,a = \alpha^{*}\,\bra{\alpha}
\end{equation}
There is no need for $\alpha$ to be a real value, as the lowering operator $a$ is not a hermitian operator (as can be readily verified), as the complex conjugate of the lowering operator is nothing but the raising operator $a^{\dagger}$.\par
Moreover we see that:
\begin{equation}
    \langle{f|a\,f}\rangle = \langle{a^{\dagger}\,f|f}\rangle
\end{equation}
These states have a very peculiar property that, they actually lead to a minimum uncertainty product, and are thus, also referred to as the "most classically behaving" states.

\vspace{5mm}
\fbox{\begin{minipage}{27em}
Coherent States are MINIMUM UNCERTAINTY STATES.
\end{minipage}}
\smallskip
\newline

The above statement shall be proved in the following section.
% \footnote{Prob 3.40 Griffiths(2^{nd} Ed.)}

\subsubsection{Minimum Uncertainty States}

We assert as of now, that the coherent state $\ket{\alpha}$ is normalized.(Later, we will determine the normalization constant exactly). That is:
\begin{equation}
    \bra{\alpha}\ket{\alpha} = 1
\end{equation}
Our goal is to determine the uncertainty product of the position operator and momentum operator of the coherent state $(\sigma_x \sigma_p)$.
The standard deviation of an arbitrary operator \^c is given as:

\begin{equation}
    \sigma_c = \sqrt{\langle{c^2}\rangle - {\langle{c}\rangle}^2}
\end{equation}
Hence we need to determine the values $\langle{x}\rangle ,\langle{x^2}\rangle,\langle{p}\rangle$ and $\langle{p^2}\rangle$.

\begin{equation}
    \langle{x}\rangle =\sqrt{\frac{\hbar}{2m\omega}} \bra{\alpha} \ket{(a + a^{\dagger})\alpha}
\end{equation}

\begin{equation}
    \langle{x}\rangle = \sqrt{\frac{\hbar}{2m\omega}}[ \bra{\alpha}\ket{a\,\alpha} + \bra{\alpha}\ket{a^{\dagger}\,\alpha}]
\end{equation}
\newline
From (1.1.3) we can readily the replace the second term as follows:
\begin{equation}
    \langle{x}\rangle = \sqrt{\frac{\hbar}{2m\omega}} [\bra{\alpha}\ket{a\,\alpha} + \bra{a\,\alpha}\ket{\alpha}]
\end{equation}
\begin{equation}
    \langle{x}\rangle = \sqrt{\frac{\hbar}{2m\omega}} [\alpha \bra{\alpha}\ket{\alpha} + \alpha^{*} \bra{\alpha}\ket{\alpha}]
\end{equation}
And so, we finally have:

\begin{equation}
    \langle{x}\rangle = \sqrt{\frac{\hbar}{2m\omega}}(\alpha + \alpha^{*})
\end{equation}
Next, to find $\langle{x^2}\rangle$:
\begin{equation}
    \langle{x^2}\rangle = \frac{\hbar}{2m\omega} \bra{\alpha}\ket{(a + a^{\dagger})^2 \alpha}
\end{equation}

    \begin{equation}
    \langle{x^2}\rangle = \frac{\hbar}{2m\omega} \bra{\alpha}\ket{(a^2 + a^{ \dagger 2} + aa^{\dagger} + a^{\dagger}a) \alpha}
\end{equation}

\begin{equation}
        \langle{x^2}\rangle = \frac{\hbar}{2m\omega}[ \bra{\alpha}\ket{a^2\,\alpha} + \bra{\alpha}\ket{a^{ \dagger 2}\,\alpha} + \bra{\alpha}\ket{ aa^{\dagger}\, \alpha} + \bra{\alpha}\ket{a^{\dagger}a\,\alpha}]
\end{equation}

Remembering that,

$$aa^{\dagger} = a^{\dagger}a + 1  $$

We can replace the operator in the third term by the R.H.S of the above expression. We therefore get,

\begin{equation}
\langle{x^2}\rangle = \frac{\hbar}{2m\omega}[ \bra{\alpha}\ket{a^2\,\alpha} + \bra{\alpha}\ket{a^{ \dagger 2}\,\alpha} + \bra{\alpha}\ket{ (a^{\dagger}a + 1)\, \alpha} + \bra{\alpha}\ket{a^{\dagger}a\,\alpha}]
\end{equation}

\begin{equation}
\langle{x^2}\rangle = \frac{\hbar}{2m\omega}[ {\alpha}^2 \bra{\alpha}\ket{\alpha} + {\alpha}^{*2}\bra{\alpha}\ket{\alpha} + (\alpha \alpha^{*} + 1) \bra{\alpha}\ket{\alpha} + \alpha\alpha^{*} \bra{\alpha}\ket{\alpha}]
\end{equation}

\begin{equation}
\langle{x^2}\rangle = \frac{\hbar}{2m\omega}( {\alpha}^2 + {\alpha}^{*2} + 2|\alpha^{2}| + 1 )
\end{equation}

Now, we can compute the standard deviation of the position operator:
\begin{equation}
\sigma_x = \sqrt{\langle{x^2}\rangle - {\langle{x}\rangle}^2}
\end{equation}

\begin{equation}
 \sigma_x = \sqrt{\frac{\hbar}{2m\omega}( {\alpha}^2 + {\alpha}^{*2} + 2|\alpha^{2}| + 1 ) - \left[{\sqrt{\frac{\hbar}{2m\omega}}(\alpha + \alpha^{*})}\right] ^2 }   
\end{equation}

\begin{equation}
    \sigma_x = \sqrt{\frac{\hbar}{2m\omega}} \sqrt{{\alpha}^2 + {\alpha}^{*2} + 2|\alpha^{2}| + 1 - ({\alpha}^2 + {\alpha}^{\dagger 2}} + 2{|\alpha|}^2)
\end{equation}
Which will give us,
\begin{equation}
\sigma_x = \sqrt\frac{\hbar}{2m\omega}
\end{equation}
\newline
Similarly, we repeat the same process to obtain $\langle{p}\rangle$, $\langle{p^2}\rangle$.

\begin{equation}
    \langle{p}\rangle = \frac{1}{i} \sqrt{\frac{m\omega \hbar}{2}} (\alpha - {\alpha}^{*})
\end{equation}

\begin{equation}
        \langle{p^2}\rangle = - \frac{m\omega \hbar}{2} ({\alpha}^2 + {\alpha}^{*2} - 2{|\alpha|}^2 -1)
\end{equation}

Which will give us,
\begin{equation}
\sigma_p = \sqrt\frac{m\omega \hbar}{2}
\end{equation}

The uncertainty product is:
\begin{equation}
    \sigma_x \sigma_p = \sqrt{\frac{\hbar}{2m\omega}} \times \sqrt{\frac{m\omega \hbar}{2}}=\frac{\hbar}{2}
\end{equation}

If we were to calculate the uncertainty product for some arbitrary energy eigen-state $\ket{n}$, we would get:

\begin{equation}
    \sigma_x \sigma_p = \frac{\hbar}{2} (2n + 1)^2
\end{equation}

Thus, we see that when $n = 0$, the uncertainty product is the least, and hence, $\ket{0}$ is also a coherent state. In the next section, we attempt to normalize the coherent state, and obtain its normalization constants.

\subsubsection{Normalization Constants Of Coherent States.}

Let us initially set the eigenvalues of the coherent states to be $c_n$ where $n$ corresponds to the energy eigen-state $\ket{n}$

\begin{equation}
  \ket{\alpha} = \sum_{n=0}^{\infty} c_n \ket{n}  
\end{equation}

The co-efficients $c_n$ are determined by the inner product of the respective stationary state $n$ and the coherent state $\alpha$

\begin{equation}
    c_n = \bra{n}\ket{\alpha}
\end{equation}

We know that:
\begin{equation}
    (a^{\dagger})^n \ket{0} = \sqrt{n!} \ket{n}
\end{equation}
And so,
\begin{equation}
    \bra{n}=\frac{1}{\sqrt{n!}}\bra{0 \quad (a^{\dagger})^n} \ket{\alpha}
\end{equation}
and the same implies also for the eigen ket states. Substituting this in $eqn(3.3.2)$ we get:

\begin{equation}
c_n = \frac{1}{\sqrt{n!}} \bra{0 \quad (a^{\dagger})^n} \ket{\alpha}     
\end{equation}

The operator $(a^{\dagger})^n$ can be made to act on the $\ket{\alpha}$  by taking its complex conjugate.We would like to remember here that $(a^{\dagger})^* = a$
\begin{equation}
    c_n = \frac{1}{\sqrt{n!}} \bra{0}\ket{(a)^n \alpha}
\end{equation}

\begin{equation}
    c_n = \frac{\alpha^n}{\sqrt{n!}} \bra{0}\ket{\alpha} 
\end{equation}
We do not yet know the inner product of the ground state and the coherent state so we will call it $c_0$.

\begin{equation}
    c_n = \frac{\alpha^n}{\sqrt{n!}} c_0
\end{equation}
By using the fact that the sum of all the probabilities of every corresponding state should be 1, that is , $\sum^{\infty}_{n=0} |c_n|^2 = 1$, we can obtain the value of $c_0$.
\begin{equation}
    |c_0|^2 \sum^{\infty}_{n=0} \frac{|\alpha|^{2n}}{n!} = 1
\end{equation}
Recalling that $e^x = \sum^{\infty}_{n=0}\frac{x^n}{n!}$
\begin{equation}
    |c_0|^2 e^{|\alpha|^2} =1    
\end{equation}
And finally we obtain $c_0$ as:
\begin{equation}
    c_0 = e^{- \frac{|\alpha|^2}{2}}
\end{equation}
Thus, the coherent state is written (in all its glory) as:
\begin{equation}
    \ket{\alpha} = e^{- \frac{|\alpha|^2}{2}} \sum_{n=0}^{\infty} \frac{\alpha^n}{\sqrt{n!}} \ket{n}
\end{equation}

Let us check by operating the above coherent state by the lowering operator $a$, whether $a$ really gives us eigen values of the state $\ket{\alpha}$:

\begin{equation}
a \ket{\alpha} = e^{-\frac{|\alpha|^2}{2}} \sum_{n=0}^{\infty} \frac{\alpha^n}{\sqrt{n!}} a\ket{n}
\end{equation}

\begin{equation}
    a\ket{\alpha} = e^{\frac{-|\alpha|^2}{2}} \left\{ \frac{\alpha^0}{\sqrt{0!}} a\ket{0} + \frac{\alpha^1}{\sqrt{1!}} a\ket{1} + \frac{\alpha^2}{\sqrt{2!}} a\ket{2} + \frac{\alpha^3}{\sqrt{3!}} a\ket{3}+... \right\}
\end{equation}

\begin{equation}
    a\ket{\alpha} = e^{-\frac{|\alpha|^2}{2}} \left\{ 0 + \frac{\alpha}{1} \sqrt{1} \ket{0} + \frac{\alpha^2}{\sqrt{2!}} \sqrt{2}\ket{1}+ \frac{\alpha^3}{\sqrt{3!}} \sqrt {3}\ket{2} +... \right\}
\end{equation}

Taking $\alpha$ common outside:

\begin{equation}
    a\ket{\alpha} = \alpha \cdot e^{-\frac{|\alpha|^2}{2}} \left\{ \ket{0} + \frac{\alpha}{\sqrt{1\cdot 2}} 2 \ket{1} + \frac{\alpha^2}{\sqrt{1\cdot 2\cdot 3}} 3\ket{2} + ...  \right\}
\end{equation}

\begin{equation}
    a\ket{\alpha} = \alpha \cdot e^{-\frac{|\alpha|^2}{2}} \left\{ \ket{0} + \frac{\alpha}{\sqrt{1!}} \ket{1} + \frac{\alpha^2}{\sqrt{2!}} \ket{2} +... \right\}
\end{equation}

Which gives us:
\begin{equation}
a\ket{\alpha} = \alpha \cdot e^{-\frac{|\alpha|^2}{2}}  \sum_{n=0}^{\infty} \frac{\alpha^n}{\sqrt{n!}} \ket{n} = \alpha \ket{\alpha}
\end{equation}

As we see,the co-efficients of the energy eigen states contribute greatly to the "quasi classical" behavior of the coherent state. The following plot shows us the absolute squares of the co-efficients of the first few stationary states, for different values of $|\alpha|^2$, \textbf{which shall be designated as $\nu$}.

\begin{figure}[H]
    \centering
    \includegraphics[width=15cm]{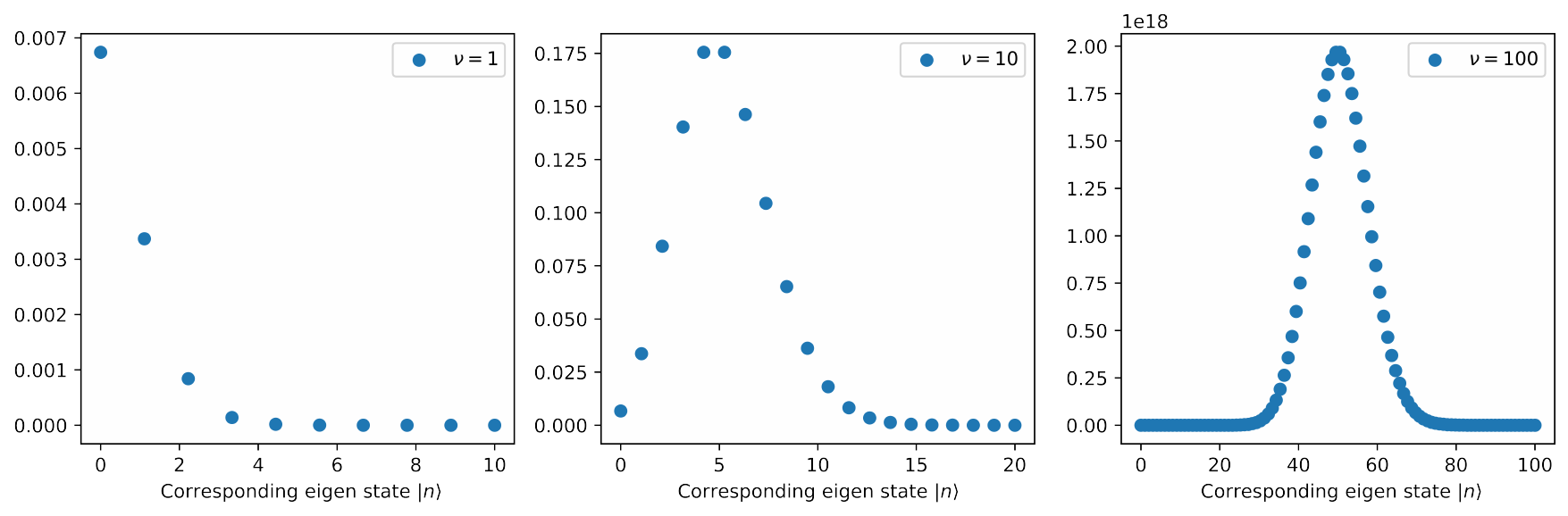}
    \label{fig:acr}
    \caption{The absolute squares of the co-efficients of coherent states with for i) $\nu = 1$, ii) $\nu = 10$, iii)$\nu = 100$ for $H = (a^{\dagger}a+\frac{1}{2})\hbar \chi$.}
\end{figure}

Hence we see that as the value of $\alpha$ increases, the co-efficients take on a more gaussian form with the maximum value at the point $\frac{|\alpha|^2}{2}$.

We also notice that, in terms of the raising operator $a^{\dagger}$, the coherent state can be represented as:
\begin{equation}
    \ket{\alpha} = e^{-\frac{|\alpha|^2}{2}} \sum^{\infty}_{n=0} \frac{(\alpha a^{\dagger})^n}{n!} \ket{0} 
\end{equation}

From equation(1.3.37) we can compute the value $|\alpha|^2$ (which we know is unity) as follows:

\begin{equation}
    |\alpha|^2 =\braket{\alpha}{\alpha} = e^{-|\alpha|^2} \sum_{m,n=0}^\infty \frac{\alpha^{* m} \alpha^n}{\sqrt{m!} \sqrt{n!}} \braket{m}{n}  
\end{equation}

\begin{equation}
    \braket{\alpha}{\alpha} = e^{-|\alpha|^2} \sum_{n=0}^\infty \frac{|\alpha|^2}{n!} = 1
\end{equation}

The co-efficients $|c_n|^2$ corresponding to each stationary state are: $|c_n|^2 = e^{-|\alpha|^2} \frac{|\alpha|^2}{n!}$ which correspond to a poisson distribution whose mean value and variance is equal to $|\alpha|^2$.

The most interesting feature perhaps, about these coherent states is that the simple harmonic potential ($ V(x) \propto x^2 $) preserves the "shape" of the wave-packet comprising of the superposition of these stationary eigen-states. Unfortunately, the same cannot be said for potentials which are not simple harmonic in nature.

\subsubsection{An interesting phenomenon}

The following is an intuitive way of understanding the almost classical behavior of coherent states.
Let's try to calculate the expression $\langle{x}\rangle$ of a quantum harmonic oscillator in some arbitrary energy state $n$.
\begin{equation}
    \langle{x}\rangle = \bra{n}x\ket{n}
\end{equation}

\begin{equation}
    \langle{x}\rangle = \sqrt{\frac{m \omega}{2\hbar}} [\bra{n}a\ket{n} + \bra{n}a^{\dagger}\ket{n}] 
\end{equation}

\begin{equation}
\langle{x}\rangle = \sqrt{\frac{m \omega}{2\hbar}} [\bra{n}\ket{n-1} + \bra{n}\ket{n+1}]    
\end{equation}
And since the energy eigen-states are ortho-normal(i.e., $\delta_{n,n-1} = 0$, $\delta_{n,n+1} =0$), we get:
\begin{equation}
    {\langle{x}\rangle}_{oscillator} = 0
\end{equation}
Next, we evaluate the expression ${\langle{x}\rangle}_{coherent}$:

\begin{equation}
    {\langle x \rangle}_{coherent} = \sqrt{\frac{m \omega}{2\hbar}} \bra{\alpha(t)} (a + a^{\dagger}) \ket{\alpha(t)} 
\end{equation}

Here, we would like to establish that $\alpha(t)$ denotes the coherent state, along with the time evolution, which as mentioned in a previous section can be taken to be, $e^{-i\omega t}$, and so, $\alpha(t) = \alpha e^{-i\omega t}$.
\begin{equation}
    {\langle x \rangle}_{coherent} = \sqrt{\frac{m \omega}{2\hbar}} (\alpha + \alpha^*)
\end{equation}
But $\alpha$, in itself being a complex value, can be written as $\alpha = |\alpha| e^{i\theta}$, and so $\alpha(t) = |\alpha| e^{- i(\omega t - \theta)}$ .We can rewrite the above expression as:
\begin{equation}
    {\langle x\rangle}_{coherent} = \sqrt{\frac{m \omega}{2\hbar}} 2 \Re{\alpha} = \sqrt{\frac{2m \omega}{\hbar}} |\alpha| cos(\omega t - \theta) 
\end{equation}

This is very much similar to the solution one would obtain for a classical oscillator!

Next, we shall look into two state systems, whose applications are vast in the field of Quantum Computing and Quantum Information.

\subsection{Bloch Sphere}
\subsubsection{Initial derivation}
Two state quantum systems which usually consist of a $\ket{g}$ state (ground state) (This is not the case always. The condition here is that the two states must be distinct and orthonormal.) and a $\ket{e}$ state (excited state) can be easily visualized by the BLOCH SPHERE.
Let's call the ground state as $\ket{0}$ and the excited state as $\ket{1}$. Their probability amplitudes are the values $\alpha$ and $\beta$ respectively.

\begin{equation}
    \ket{\psi} = \alpha \ket{0} + \beta \ket{1}
\end{equation}
Before observing this system, it is asserted that the system is in a superposition of the $\ket{0}$ state and $\ket{1}$ state.
\par
The very act of observing the system, forces it into one of the two possible states. The probability of the system(after measurement) to be in the $\ket{0}$ is $|{\alpha}|^2$ and for it to be in the $\ket{1}$ is $|{\beta}|^2$.

To get into the "Bloch Representation", we need to first sort out the following: 
\begin{enumerate}
    \item { Any complex number, say $\mathbb{Z}$ can be represented as a function of it's magnitude $r$ and phase $\theta$ as}:
    \begin{equation}
        \mathbb{Z} = re^{i\theta}
    \end{equation}
where, $r$ gives us the distance from the origin, and $\theta$ the angle we need to rotate from the x-axis to reach the complex number. It might as well be written as $\mathbb{Z} = p+iq$, where p and q correspond to the real and imaginary components of the complex number.
    \item {We shall represent $\alpha$ and $\beta$ as the following:}
    \begin{equation}
        \alpha = r_{\alpha} e^{i{\phi}_0} ;\quad \beta = r_{\beta} e^{i{\phi}_1}
    \end{equation}
    \item The states $\ket{0}$ and $\ket{1}$ are orthonormal to each other and cannot be represented by any linear combination of the other. 
    That is,
    \begin{equation}
        \bra{0}\ket{1} = 0
    \end{equation}
    \item The state $\ket{\psi}$ is normalized,i.e., $\bra{\psi}\ket{\psi} = 1$, so we now encounter an constraint that:
    \begin{equation}
        |\alpha|^2 +|\beta|^2 =1
    \end{equation}
    which, we realize, bears much resemblance to the trigonometric identity $cos^2 \theta + sin^2 \theta = 1$. 
\end{enumerate}
So, as of yet, let's naively assert the following:
\begin{equation}
 |\alpha|^2 = cos^2 \theta ; \quad |\beta|^2 = sin^2 \theta   
\end{equation}
From which we can rewrite state $\ket{\psi}$ as:
\begin{equation}
    \ket{\psi} = cos(\theta) e^{i{\phi}_0} \ket{0} + sin(\theta) e^{i{\phi}_1} \ket{1}
\end{equation}
Let's take the term $e^{i{\phi}_0}$ common:
\begin{equation}
   \ket{\psi} = e^{i{\phi}_0} (cos(\theta) \ket{0} + sin(\theta) e^{i({\phi}_1 - {\phi}_0)}\ket{1}) 
\end{equation}
Let's set ${\phi}_1 - {\phi}_0 = \phi$

As $\psi$ is normalized, the overall phase $e^{i{\phi}_0}$ is of no consequence here, and can be neglected without any loss of generality.

So, we see that we can write $\ket{\psi}$ as:
\begin{equation}
    \ket{\psi} = cos(\theta) \ket{0} + \sin(\theta) e^{i{\phi}} \ket{1}
\end{equation}

This is similar to the spherical polar representation of a unit vector:
\begin{equation}
    \hat{r} = sin(\theta) cos(\phi) \hat{x} + sin(\theta) sin(\phi) \hat{y} + cos(\theta) \hat{z}
\end{equation}
So, it is evident, that we are approaching a spherical form of representation of the system

But then again, we encounter a problem!
\subsubsection{Half Angles}
We know that, $\theta$ runs from 0 to $\pi$ ,and $\phi$ from 0 to $2\pi$. When $\theta = 0$, we get:
\begin{equation}
   \ket{\psi} = cos(0) \ket{0} + sin(0) e^{i\phi} \ket{1} = \ket{0} 
\end{equation}
And when $\theta = \pi$:
\begin{equation}
    \ket{\psi} = cos(\pi) + sin(\pi) e^{i\phi} \ket{1} = -\ket{0}
\end{equation}
But, then again, we have no need to represent a system with negative probability amplitudes, because upon normalization, they will yield the same probabilities as their positive counterparts. So, we are only concerned with one of the two hemispheres. For ease of calculation, we will only consider the upper hemisphere(while the lower hemisphere is also equally valid here).

So, we might as well confine ourselves to the range $\theta = 0$ to $\frac{\pi}{2}$. In our new co-ordinate system, we have thus found that:
\begin{equation}
    \theta = 2\theta' \rightarrow \theta' = \frac{\theta}{2} 
\end{equation}
 And so, finally, we can represent $\ket{\psi}$ as:
 \begin{equation}
     \ket{\psi} = cos(\frac{\theta}{2}) \ket{0} + sin(\frac{\theta}{2}) e^{i\phi} \ket{1}
 \end{equation}

A typical Bloch sphere looks like this:
\begin{figure}[H]
    \centering
    \includegraphics[width=6cm]{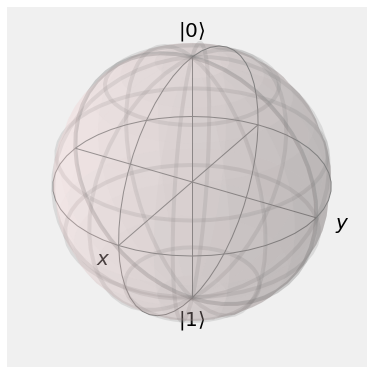}
    \label{fig:bloch}
\end{figure}

Let's plot a state onto the sphere.

For instance, let's try to plot a state which has equal probability of being in the $\ket{0}$ state and the $\ket{1}$ state:

\begin{equation}
    \ket{\psi} = \frac{1}{\sqrt{2}} \ket{0} + \frac{1}{\sqrt{2}} \ket{1}
\end{equation}

To do so, we find that we need to set, $\theta = \frac{\pi}{2}$ and $\phi =0$.
\begin{figure}[H]
    \centering
    \includegraphics[width=6cm]{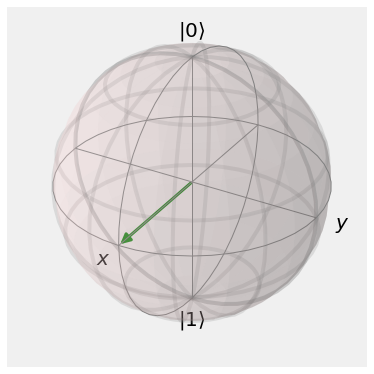}
    \label{fig:bloch}
\end{figure}
The unit vector from the origin to the point on the surface of the Bloch sphere is called the \textit{Bloch Vector}. 
The above state is commonly called as the $\ket{+}$ state.

\subsubsection{Two State Oscillator}

Let us consider a two-state system which evolves in time with accordance to the Hamiltonian:

\begin{equation}
    \hat{H} = \hbar \omega (a^{\dagger}a + \frac{1}{2})
\end{equation}

The state $\psi$ as defined previously along with the Time Evolution Operator is given as:

\begin{equation}
    \ket{\psi} = (\alpha \ket{0} + \beta \ket{1}) e^{-i \frac{\hat{H}t}{\hbar}} 
\end{equation}

\begin{equation}
    \ket{\psi} = (\alpha e^{-i \frac{E_0 t}{\hbar}} \ket{0} + \beta e^{-i \frac{E_1 t}{\hbar}} \ket{1})
\end{equation}
which might as well be written as: (keeping in mind that the overall phase plays no role here),
\begin{equation}
    \ket{\psi} = (\alpha \ket{0} + \beta e^{i \frac{(E_0 - E_1)t}{\hbar}} \ket{1})
\end{equation}
Comparing the above equation with the standard form of a qubit, we find that the frequency of rotation of the Bloch Vector in the surface of the Bloch Sphere, $\omega'$ :
\begin{equation}
    \omega' = \frac{(E_0 - E_1)}{\hbar}
\end{equation}

The above argument can also be extended to \textbf{spin systems}, without any loss of generality.

\section{Revivals and Fractional Revivals}
% \vskip 2em
 \par Owing to the unitary transformation brought about by the time-evolution operator, the system during its evolution in time, loses its amplitude and then comes back to its original state after some periodic interval. This phenomenon can be viewed in all quantum systems when we let them evolve in time. \newline
 \par This behaviour of the quantum system is dubbed as the "revival" of the system~\cite{sudheesh_2004_frac_revivals}.
 In the further sections, we shall look into this concept in greater detail.
 
Simply put, we wish to show that at some specific time interval the probability of finding the state in the exact same configuration is equal to what it was initially.

Let's consider an elementary quantum system (in some detail) and work out its revival time to understand this phenomenon better.
 
Consider the wave function of an particle in an infinite square well~\cite{styer_quantum_revivals_2001} of width a.\footnote{Prob 2.41 Griffiths ($2^{nd}$ Ed.)} Is there a time $t= T_{rev}\geq 0$, for which the wave-function $\psi(x, T_{rev})$, becomes equal to $\psi(x,t = 0)$? 
\newline To solve this, we need to consider the general solution of the time-dependent Schrodinger equation of a particle in an infinite well.(which has been solved by us in an extensive manner, as part of the coursework.)
\begin{equation}
    \psi(x,t) = \sum_{n=1}^{\infty} c_n \psi_n(x) e^{-i \frac{n^2 \pi^2 \hbar}{2ma^2}t }
\end{equation}

Only $\mathbb{U}(t) = e^{-i\frac{n^2 \pi^2 \hbar^2}{2ma^2}t}$ is a function of time, and so, we need to find that time $T_{rev}$, for which $\mathbb{U}(t=0) = \mathbb{U}(t = T_{rev})$. We know that for a function of form $f(x) = e^{ix}$, when $x=0, 2\pi, 4\pi..,$, the value of $f(x) = 1$, and so, we can see that, the wave-function "revives" when:
\begin{equation}
    \frac{n^2 \pi^2 \hbar }{2ma^2}T_{rev} = 2\pi
\end{equation}
 The value of $n^2$ is irrelevant here, as it will take some integral value $>1$. Hence, we can eliminate that term, and consider the remaining:
 
 \begin{equation}
     \frac{\pi^2 \hbar}{2ma^2}T_{rev} = 2\pi
 \end{equation}
 From which, we get:
 \begin{equation}
     T_{rev} = \frac{4ma^2}{\pi \hbar}
 \end{equation}
 And so, we find that the below given expression holds true:
 
 \begin{equation}
     \psi(x, t + n\left( \frac{4ma^2}{\pi \hbar}\right)) = \psi(x,t)
 \end{equation}
 In a similar manner, for all quantum systems, the wave-function "revives" itself periodically. Further interesting relations between the classical and quantum revival times have been studied\cite{styer_quantum_revivals_2001}.

\subsection{Auto-Correlation Function}
To measure the existence of these so called revivals, one must measure the
likeliness of the state of the wave-packet at some later time period, with that of the initial wave-packet, that is, the dot product between $\ket{\psi(t=0)}$ and $\ket{\psi(t=t')}$ where $t'$ is some arbitrary time. 
\newline
This function $|\bra{\psi(t)}\ket{\psi(0)}|^2$ is called as the \textbf{auto-correlation function}, and can be used to measure the temporal coherence of the wave-function as time evolves.
\newline 
Let us initially consider a state $\ket{\psi}$, whose basis vectors are $\ket{\phi_0}$, $\ket{\phi_1}$,..,$\ket{\phi_n}$, and the corresponding probability amplitudes associated with each basis vector is given by $c_0$, $c_1$ ,.., $c_n$, and whose time evolution is governed by an Hamiltonian $\hat{H}$.
We see that at $t=0$, the state $\psi$ is given as:
\begin{equation}
    \ket{\psi(0)} = \sum_{n=0}^{\infty} c_n e^{-\frac{i\hat{H}(0)}{\hbar}} \ket{\phi_n} 
\end{equation}

\begin{equation}
    = \sum_{n=0}^{\infty} c_n \ket{\phi_n}
\end{equation}

At some arbitrary time t, the state $\psi$ is given as:
\begin{equation}
    \bra{\psi(t)} = \sum_{n=0}^{\infty} c^*_n e^{-\frac{i\hat{H}t}{\hbar}} \bra{\phi_n} 
\end{equation}

The inner product of $\psi(t)$ and $\psi(0)$ can be written as:

\begin{equation}
    \braket{\psi(t)}{\psi(0)} = \sum_{n=0}^{\infty} |c_n|^2 e^{-\frac{i\hat{H}t}{\hbar}} 
\end{equation}

The appropriate normalization constants and Hamiltonian can be provided to the above relation to evaluate the auto-correlation function of the quantum system in question. 

In the following sections, we shall look into certain Hamiltonians and plot their auto-correlation functions.
In other words, the auto-correlation function provides us the measure of likeliness of the state at initial time and the state as time evolves. 
When this function returns to the initial value at some later instant of time, say at $t=t'$, we say that the state has "revived" to its initial condition and this phenomenon is called "quantum revival". 
\newline \par In the case of the harmonic oscillator, the hamiltonian is $H = (a^{\dagger}a + \frac{1}{2}) \hbar \chi$ , and hence the auto-correlation function looks like:
\begin{equation}
    \braket{\psi(t)}{\psi(0)} = e^{-|\alpha|^2} \sum_{n=0}^{\infty} \frac{|\alpha|^{2n}}{n!} e^{in\chi t}
\end{equation}

\begin{figure}[H]
    \centering
    \includegraphics[width=15cm]{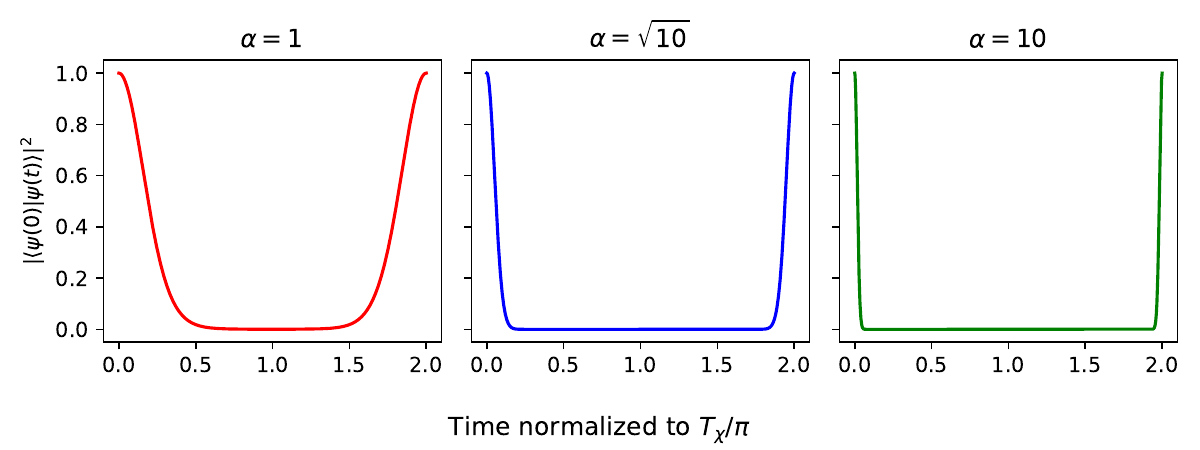}
    \label{fig:acr}
    \caption{Absolute square of Auto Correlation Function for i) $\nu = 1$, ii) $\nu = 10$, iii)$\nu = 100$ for $H = (a^{\dagger}a+\frac{1}{2})\hbar \chi$.}
\end{figure}

As we can see, the wave-function smoothly varies as it loses its resemblance, and at half the revival time, the wave-function is the farthest away from the initial state, and slowly it comes back to its original state at the revival time. 

\subsubsection{Fractional Revivals}

The basic premise of fractional revivals is that, at any given moment after the propagation of a quantum wave packet (of a coherent state) in a non-linear medium (or whose time evolution is guided by a non-linear Hamiltonian), the wave packet becomes a superposition of several coherent states,wherein the exact amount of these coherent states is determined by the ratio between that specific instant of time and that of the time taken for full revival.
\par We shall look into a quantum system in general, and investigate\cite{styer_quantum_revivals_2001}\cite{silva2023revivalpatternsdiraccat}\cite{Rohith_2024}\cite{PhysRevA.90.012320} why these so-called "fractional revivals" happen. Any quantum particle or group of quantum particles, which are described by a single wave-function can be written in terms of the ortho-normal energy eigen-states $\phi_n$.

\begin{equation}
  \psi(t) = \sum_{n=0}^\infty c_n \phi_n e^{-i E_n t} 
\end{equation}

The auto-correlation function of the function $\psi(t)$ is:

\begin{equation}
    \braket{\psi(t)}{\psi(0)} = \sum_{n=0}^{\infty} |c_n|^2 e^{iE_n t}
\end{equation}

As $n \longrightarrow \infty$ the expansion co-efficients will take on a gaussian form and their maximum will be at some point, say $n=n_0$, and $|c_n|^2$ can be written of form:

\begin{equation}
    |c_n|^2 = \frac{e^{-(n-n_0)^2 / 2\sigma^2}}{\sqrt{2\pi \sigma}}
\end{equation}
where $\sigma$ is nothing but the width of the gaussian wave-packet or function.
The assumption we are making right now is that, the energy eigenvalues $E_n$ are regular functions of n, and therefore have a Taylor series expansion about some chosen value $n=n_0$:
\begin{equation}
    E_n = E_{n_0} + {E'}_{n_0} (n-n_0) + \frac{1}{2} E''_{n_0} (n-n_0)^2 +...
\end{equation}

The co-efficient of the variable $(n-n_0)$ is the term which takes care of the propagation of the wave, or in other words the term which decides the revival time. As seen by us previously, the co-efficient of the linear term contributes to the group velocity of the individual waves constituting a superposition of infinite sinusoidal waves.

\par The co-efficient of the quadratic term $(n-n_0)^2$, or the second derivative of the energy eigenvalues with respect to n, determines the \textbf{amount of dispersion} the wavepacket undergoes. \textbf{The dispersion of a wave-packet leads to the phenomenon of "fractional revivals"\cite{fractional_revivals_universality}}.
\subsection{Non-linear Kerr Hamiltonian}
As has been stated earlier, fractional revivals are observed in a system which are described by a non-linear Hamiltonian. One such Hamiltonian\cite{griffiths_introduction_quantum_mechanics_2ed} which effectively describes the Kerr non-linearity~\cite{kerr_optomech_effective_hamiltonian} is given as the following: 
\begin{equation}
H = (a^{\dagger 2} a^2)\hbar \chi 
\end{equation}

where $\chi$ is some positive constant, $a$ and $a^{\dagger}$ are the annihilation and creation operators, and $a^{\dagger} a = N$ , and the number operator. and thus the  non-linear Hamiltonian can the precisely written in the form of:
\begin{equation}
H = \hat{N}(\hat{N}-1)\hbar \chi
\end{equation}

The conversion of expressing in the form of the raising and lowering operators, to the number operator, is done in the following way, by remembering that $[a, a^{\dagger}] = 1$, and that $a^{\dagger}a = aa^{\dagger} + 1$:
\begin{equation}
    a^{\dagger 2}a^2 = a^{\dagger}a^{\dagger}aa=a^{\dagger}(aa^{\dagger}+1)a
    =a^{\dagger}aa^{\dagger}a+ a^{\dagger}a=(a^{\dagger}a)^2 + a^{\dagger}a = \hat{N}^2 + \hat{N}=\hat{N}(\hat{N}+1)
\end{equation}

Some materials exhibit interesting phenomena\cite{George2012Ehrenfest} when we allow light to propagate through them. While usually, the refractive index $n$ of a material is directly proportional to the intensity of the electric field($\vec{\textbf{E}}$) due to the incident light, some materials have refractive indices which are additionally proportional to the square of the electric field ($|\vec{{\textbf{E}}}|^2$). The effective physics for such phenomena can be captured by the Hamiltonian(ignoring the linear part) given in eqn(2.2.1), where $\chi$ is the third order susceptibility term.

 We shall henceforth concern ourselves with the coherent state $\ket{\alpha}$  , a suitable superposition of the energy eigen states,which is an eigen state of the lowering operator $a$ , i.e.,$(a \ket{\alpha} = \alpha \ket{\alpha})$ 
The initial state at t = 0 is given as:
\begin{equation}
\ket{\alpha (t = 0)} = e^{-|\alpha|^2 / 2} \sum_{n = 0}^\infty \frac{\alpha^n}{\sqrt{n!}} \ket{n}  
\end{equation}
Adding in the time dependence, $\ket{\alpha (t)}$ is given as:
\begin{equation}
\ket{\alpha (t)} = e^{-|\alpha|^2 / 2} \sum_{n = 0}^\infty \frac{\alpha ^n}{\sqrt{n!}}e^{-i \chi N(N-1)t} \ket{n}
\end{equation}
At times, $t = T_{rev} = \frac{\pi}{\chi}$, the system returns back to it's original state, and thus revival occurs, and fractional revivals occur at times $t = \frac{\pi l}{m\chi}$
where, m = 2,3,4... and l = 1,2,...,(m-1) for a given value of m.

By fractional revivals, we mean that the gaussian wavepacket gets split up into smaller copies of itself, the number of such copies being dependent on the value of m. This can be thought of as a superposition of $n$ coherent states at the $(n-1)^{th}$ fractional revival, along with suitable normalization constants.

Hence, it shall be suffice to show that the above statement is true in the case of fractional revivals.

Here , the eigen value $\alpha$ can be written in the forms:
\begin{equation}
i) \alpha = \frac{p_1 + iq_1}{\sqrt{2}}
\end{equation}
\begin{equation}
ii)\alpha = r_1 e^{i\theta_1}
\end{equation}

where $p_1$,$q_1$ are related with $r_1$ and $\theta_1$ in the following way:
\begin{equation}
p_1 = r_1cos\theta_1;\quad q_1 = r_1sin\theta_1 
\end{equation}
Both of these definitions of $\alpha$ will be used here based on the problem at hand. 
The general formula\textbf{[1]} to obtain $\langle{a^{\dagger k} a^{k+l}}\rangle$  is given as follows:
\begin{equation}
\langle a^{\dagger k} a^{k+l}\rangle = \alpha^l |\alpha|^{2k} e^{-|\alpha|^2  (1 - cos2\chi lt)} exp[-i\chi (l(l-1)+2kl)t - i|\alpha|^2 sin2l\chi t ] 
\end{equation}

The derivation of the above expression is provided in Appendix A. 
\subsection{Auto-Correlation Function of a coherent state in a non-linear Kerr Medium}
We'd like to take a small detour at this point, and venture to plot the auto-correlation function of a coherent state whose time evolution is governed by the non-linear term of the Kerr Hamiltonian $H = a^{\dagger 2}a^2 \hbar \chi$.
\begin{equation}
    \braket{\alpha(t)}{\alpha(0)} = e^{-|\alpha|^2} \sum^{\infty}_{m,n=0} \frac{\alpha^{* m}\alpha^n}{\sqrt{m!n!}} e^{-i(n(n-1))\chi t} \braket{m}{n}
\end{equation}

\begin{equation}
    \braket{\alpha(t)}{\alpha(0)} = e^{-|\alpha|^2} \sum^{\infty}_{n=0} \frac{|\alpha|^{2n}}{n!} e^{-i(n(n-1))\chi t} 
\end{equation}
Notice that the energy of an individual wave-function is directly proportional to the square of the energy eigen-state number $n$, which is touted to be the root of the occurrence of fractional revivals in the system.
A simple manner to plot the absolute square of auto-correlation function for coherent states whose Hamiltonians are $H=H(n)$ is provided in the Appendix B

\begin{figure}[H]
    \centering
    \includegraphics[width=15cm]{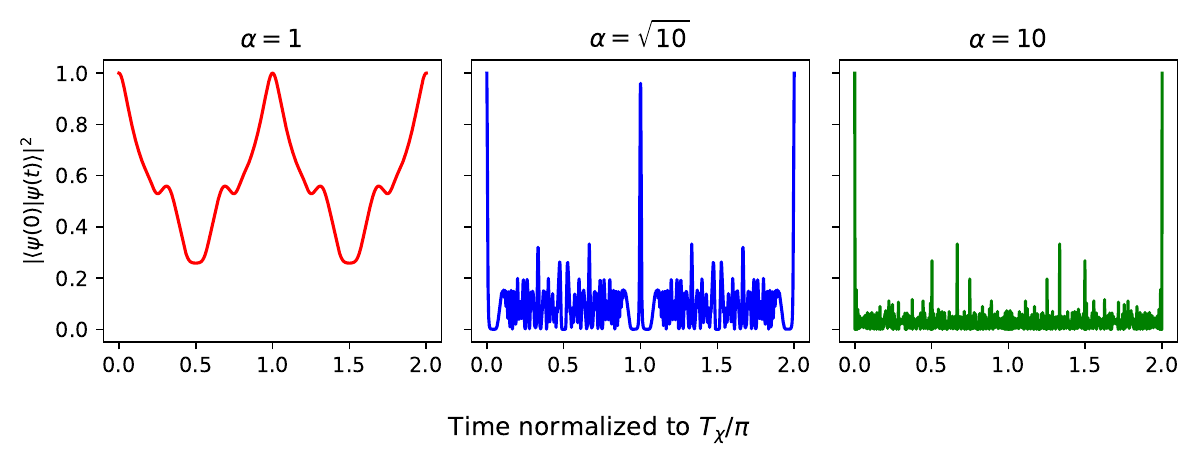}
    \label{fig:acr}
    \caption{Absolute square of Auto Correlation Function  for i) $\nu = 1$, ii) $\nu = \sqrt{10}$, iii) $\nu = 10$ for $H = a^{\dagger 2}a^2 \hbar\chi$}
\end{figure}

We see that as $\alpha$ increases, the chaotic nature of the auto-correlation function reduces, and an ordered and periodic function can be seen. This yet again, shows us that, as the value of $\alpha$ increases, the system behaves more and more like a classical system which is deterministic in nature.
\subsection{Fractional Revivals of a Coherent State in a Kerr medium}
The derivation provided in this subsection has been carried out extensively in \cite{kerr_schrodinger_cat_production}, and in a brief manner in \cite{sudheesh_nonclassical_wavepacket}.We shall confine ourselves to proving that at the instants of fractional revivals, the initial coherent state \textbf{can} be written as a superposition of coherent states, where the number of these coherent states is proportional to the corresponding fractional revival.

\par A coherent state propagating in a Kerr-medium whose Hamiltonian is of form $\chi a^{\dagger 2}a^2$ is given as:
\begin{equation}
    \ket{\alpha(t)} = e^{-i\chi \hat{N}(\hat{N}-1)t} \ket{\alpha}
\end{equation}
where, $\hat{N} = a^{\dagger}a$,the number operator. \newline

As we are interested in what happens at times which are fractions of the revival time, we shall henceforth let our time be $t= \frac{\pi}{m\chi}$, where $m$ takes only integral values greater than 1. Thus, the state-vector can now be written as:
\begin{equation}
    \ket{\alpha\left(-i\frac{\pi}{m} \hat{N}(\hat{N}-1)\right)} = e^{-i\frac{\pi}{m}\hat{N}(\hat{N}-1)}\ket{\alpha}
\end{equation}

As has been discussed previously, the term $\hat{N}^2$ leads to non-linear effects, and is the primary reason for the occurrence of fractional revivals. The time evolution operator $\mathbb{U}$ can now be written as:

\begin{equation}
    \mathbb{U} = e^{-i\frac{\pi}{m}\hat{N}(\hat{N}-1)}
\end{equation}

The values that $m$ can take are either even or odd, and these can be used by us to construct the Fourier Series of the time evolution operator. We shall proceed in the following manner, by noticing that for odd values of m, the term $\hat{N}(\hat{N}-1)$ remains as such, and for even values of m, the term transforms into $\hat{N}^2$. Considering the above statement, and making the initial conversion $\hat{N} = \hat{N}-m$ we get two individual expressions for odd and even values of $m$:

\begin{equation}
    exp\left\{ -i\frac{\pi}{m} (\hat{N}+m)(\hat{N}+m-1) \right\} = (-1)^{m-1} exp\left\{ -i\frac{\pi}{m} \hat{N}(\hat{N}-1) \right\}
\end{equation}

\begin{equation}
    exp\left\{ -i\frac{\pi}{m} (\hat{N}+m)^2 \right\} = (-1)^{m} exp\left\{ -i\frac{\pi}{m} \hat{N}^2 \right\}
\end{equation}

The equations (2.4.5) and (2.4.6) pertain to odd and even values of $m$ respectively.

The periodicity property of these functions can now be used to expand $\mathbb{U}$ as a Fourier Series\cite{kerr_schrodinger_cat_production}, with the base functions as $exp(-2\pi i q/m)$:

\begin{equation}
    exp\left\{ -i \frac{\pi}{m} \hat{N}(\hat{N}-1) \right\} = \sum_{n=0}^{m-1} f_n e^{-i\frac{2\pi n}{m}\hat{N}}
\end{equation}

\begin{equation}
    exp\left\{ -i \frac{\pi}{m} \hat{N}^2 \right\} = \sum_{n=0}^{m-1} g_n e^{-i\frac{2\pi n}{m}\hat{N}}
\end{equation}
And thus finally we get the condition:
\[
    \ket{\alpha_m}= 
\begin{cases}
    \sum_{n=0}^{m-1} f_l \ket{\alpha e^{-2\pi i n / m}}  ,& \text{if } m\, odd\\
    \sum_{n=0}^{m-1} g_l \ket{\alpha e^{i\pi m} e^{-2\pi i n /m}}  ,              & \text{if }m\, even 
\end{cases}
\]

We see that corresponding to the value of $m$, the initial coherent state can be written as a superposition of $m$ coherent states, for the values of $n=0,1,2,..,m-1$.

\subsection{Classical Analogs}

The phenomenon of quantum revivals and fractional revivals, initially seems to be very anomalous in nature and is quite difficult to comprehend. Furthermore, it has been posited by physicists to be a phenomenon which is "entirely quantum mechanical" in nature, and is a consequence of the wave-like properties exhibited by quantum particles. \newline
Thus, in this chapter, we look into a few classical systems which too exhibit similar behavior in one way or another. \newline 

\subsubsection{Talbot Effect}

Effects akin to revival and fractional revival were surprisingly observed almost two hundred years ago, in the field of optics, by Henry Fox Talbot in the year 1836. He reported that when light is allowed to pass through a grating, and the resulting diffraction pattern is observed for different distances, the result collection of such images viewed sideways together form a pattern, now famously known as "Talbot Carpet".
An image of such a Talbot Carpet is given below:
\begin{figure}[H]
  \begin{center}
    \includegraphics[width=0.48\textwidth]{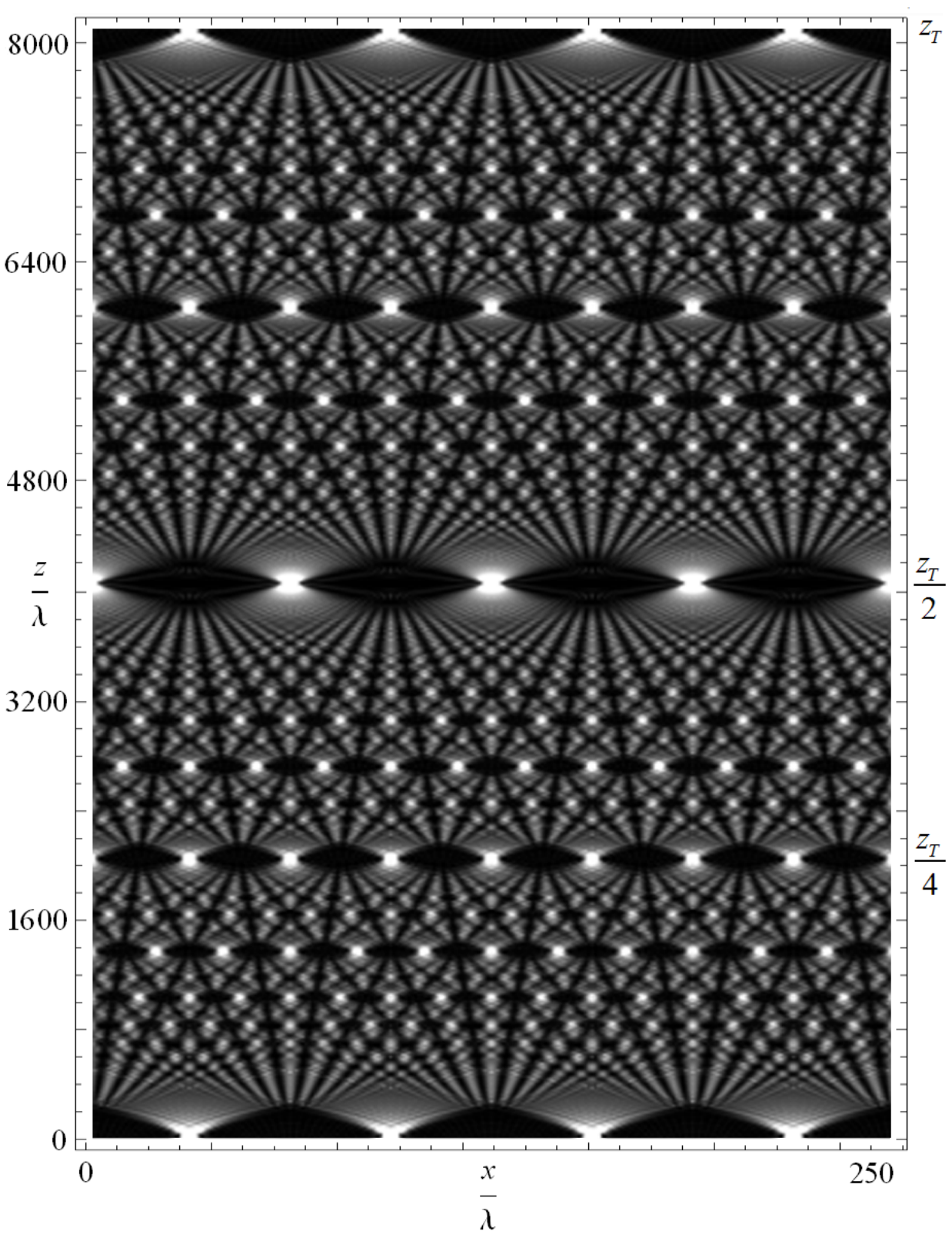}
  \end{center}
  \caption{ An Optical Talbot Carpet. \href{https://upload.wikimedia.org/wikipedia/commons/6/69/Optical_Talbot_Carpet.png}{Image Source} }
\end{figure}
The Talbot Length, or the distance after which the image repeats itself is given by Lord Rayleigh\cite{Rayleigh01031881} in the following manner:
\begin{equation}
    z_T = \dfrac{\lambda}{1- \sqrt{1-\frac{\lambda^2}{a^2}}}
\end{equation}
where,\newline
a = period of diffraction grating
$\lambda$ = wavelength of the incident light

We can see that, at half the Talbot length, we get a self image of the initial image but phase-shifted by half a period, and at one-fourth of the Talbot length, two smaller self-images of the initial image is created, similarly, for smaller and smaller intervals of the Talbot length, we see fractal like self-images. This is analogous to the phenomenon of fractional revivals. Similar to these "Talbot Carpets", there are "Quantum Carpets" in the quantum realm, about which we shall discuss briefly in section 3.5.

\subsubsection{Dancing Pendulums}

The exact origin of this arrangement is unknown, but many variations of this have been seen by us throughout the years\cite{pendulum_waves_aliasing}\cite{pendulum_waves_demonstration}.  \newline

\begin{figure}[t]
  \begin{center}
    \includegraphics[width=0.48\textwidth]{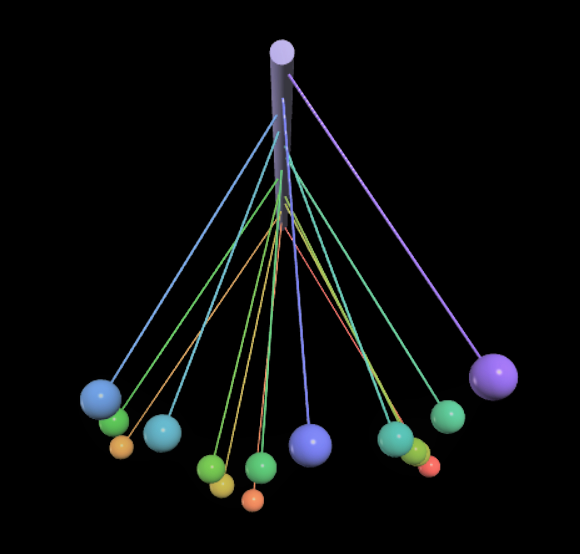}
  \end{center}
  \caption{ An apparatus of 15 pendulums, with monotonically increasing lengths. The link for this animation is provided in Appendix C}
\end{figure}

A series of uncoupled pendulums, each of varying lengths are arranged in such a manner that the time period of each succeeding pendulum is incremented by the same value.
That is:
\begin{equation}
    T_{n+1} = T_n + 1
\end{equation}

We know that in the case of a simple pendulum whose length is $l$, upon solving the lagrangian, and confining oneself to small angles ($sin \theta \approx \theta $), we get:

\begin{equation}
    \ddot{\theta} \approx - \dfrac{g}{l} \theta 
\end{equation}

whose solution is:
\begin{equation}
    \theta (t) = A cos(\omega t + \phi) + B sin(\omega t + \phi)
\end{equation}

This tells us that the frequency of oscillation, is given as: $\omega = \sqrt{\dfrac{g}{l}} $ from which we can infer that:
\begin{equation}
    T_{osc} =  2\pi \sqrt{\frac{l}{g}} 
\end{equation}

We saw earlier that when a non-linear Hamiltonian acts on a coherent state, then the time evolution operator was of form:
\begin{equation}
\mathcal{U} = e^{-i n(n-1)\chi t}
\end{equation}

And in the case of coherent states $n \longrightarrow \infty$, we might as well approximate $n(n-1) \longrightarrow n^2$. As a result of this, the wave-packet constitutes of sinusoidal waves whose frequencies increase as the square of their order. 

In the case of these dancing pendulums, similar to the quantum fractional revivals, the strength of each individual wave changes as a function of the fraction of the revival time. In the below given plot, the snapshots of 100 linear oscillators (whose spring constants are set in such a way that the consecutive oscillators, complete exactly one more oscillation than its previous one) are provided at the time intervals a)$\frac{T_{rev}}{2}$, b)$\frac{T_{rev}}{4}$ and c)$\frac{T_{rev}}{10}$.

\begin{figure}[H]
    \centering
    \includegraphics[width=15cm]{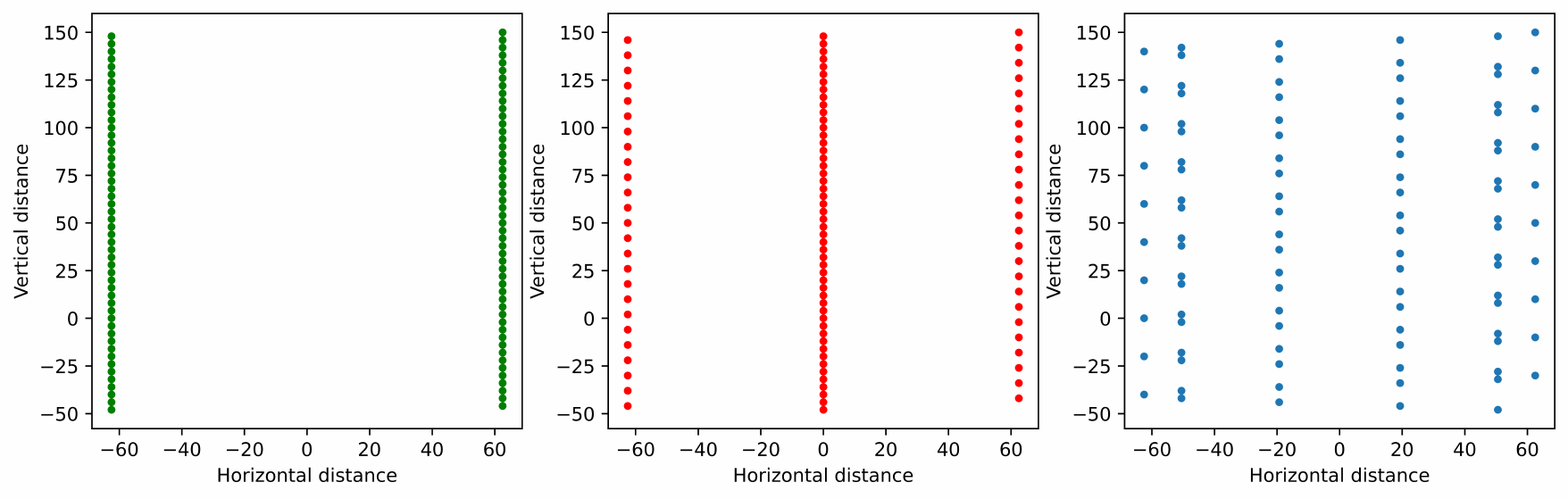}
    \label{fig:acr}
    \caption{ The position of 100 oscillators at times a)$\frac{T_{rev}}{2}$, b) $\frac{T_{rev}}{4}$, c)$\frac{T_{rev}}{10}$ }
\end{figure}

The following can be perceived from these plots.\newline
a) The initial configuration of 100 oscillators has now been separated into two "waves" each containing 50 oscillators.
\newline
\textbf{No.of oscillators in an individual wave = 50}\newline
b) At one-fourth of the revival time, the 100 oscillators have been separated into 4 individual waves, and two of these waves are at $x=0$.
\newline
\textbf{No.of oscillators in an individual wave = 25}\newline
c)At one-tenth of the revival time, there are 10 individual waves each containing 10 oscillators, eight of which are in four sub groups each containing two individual waves.
\newline
\textbf{No.of oscillators in an individual wave = 10}\newline

In this manner, we can construct a plot, which shows us the strength of individual "waves" as a function of time.
\begin{figure}[h]
    \centering
    \includegraphics[width=15cm]{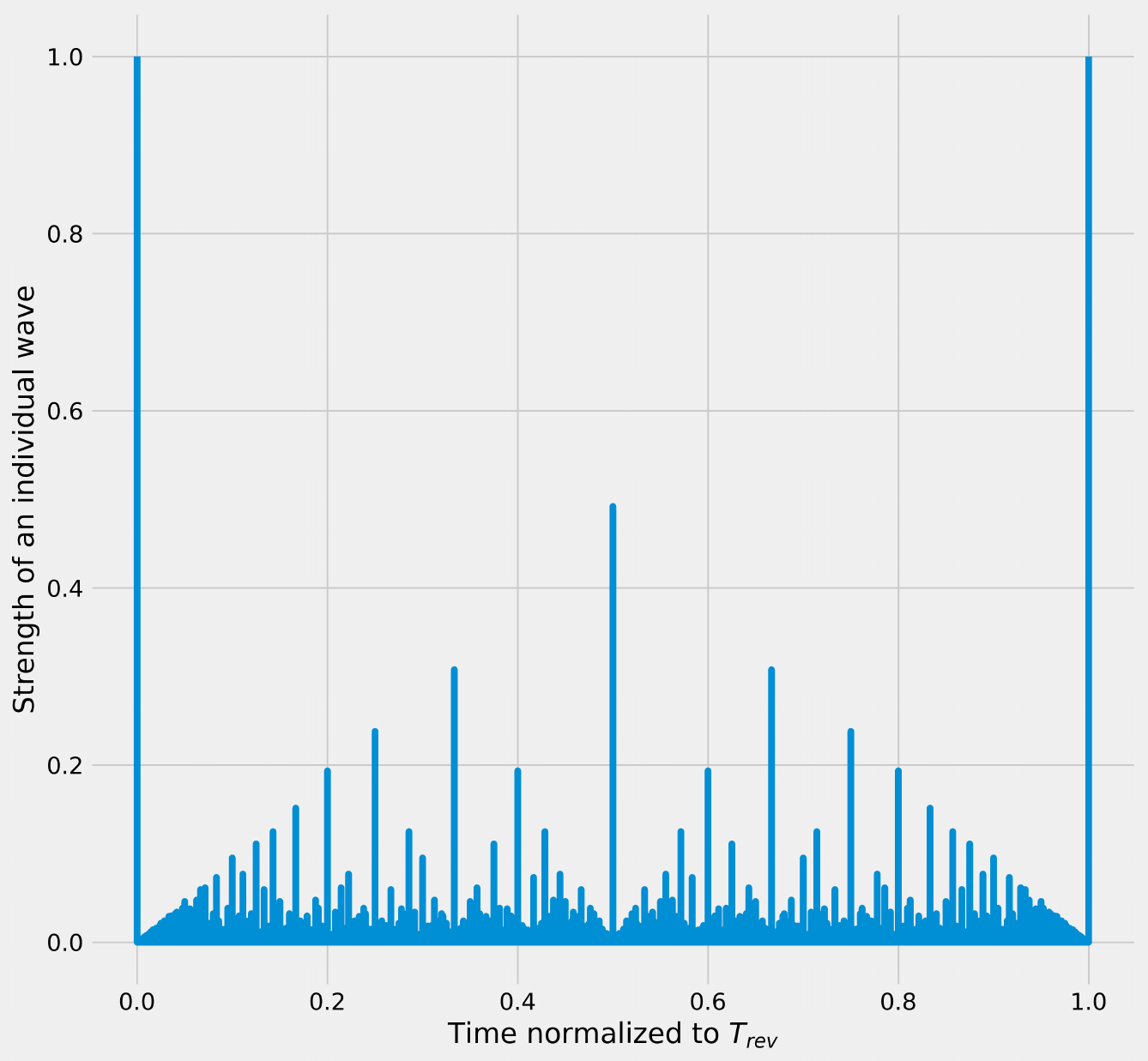}
    \label{fig:acr}
    \caption{ Strength of individual wave vs Time normalized to Full Revival Time for 50,400 oscillators. }
\end{figure}
Here, we can clearly that the strength of the individual waves, are determined by the expression $\frac{j}{k}T_{rev}$, where $k>1$ and $j=1,2,..,k-1$. This is the same expression we use to denote the fractional revivals of a quantum system.

\section{Visualizing Revivals and Fractional Revivals}

\subsection{Calculating the Expressions for Moments of Observables}
While these "fractional revivals" can be proved to exist theoretically, one does wonder, if there is any possible way for us to observe its manifestations experimentally.Signatures of these fractional revivals\cite{sudheesh_2004_frac_revivals} can be seen in observables in a limited manner, that is, one can observe the signatures of the $k^{th}$ fractional revival in the plot of the $(k+1)^{th}$ moment of observable\cite{sudheesh_nonclassical_wavepacket}. We shall rewrite the general formula to obtain the expectation value of an operator of form $a^{\dagger l} a^{k+l}$:

\begin{equation}
\langle a^{\dagger k} a^{k+l}\rangle = \alpha^l |\alpha|^{2k} e^{-|\alpha|^2  (1 - cos2\chi lt)} exp[-i\chi (l(l-1)+2kl)t - i|\alpha|^2 sin2l\chi t ] 
\end{equation}
From the above expression the following\cite{sudheesh_nonclassical_wavepacket} can be inferred:

\begin{enumerate}
    \item The sinusoidal terms in the expression control the temporal frequency of the function. The presence of a $2l\chi$ as the frequency of a sinusoidal function in the phase term, tell us that the value given to $l$ determines the frequency of the function.
    \item The term $e^{-|\alpha|^2 (1-cos2\chi l t)}$ does not affect the function when the value of $\alpha$ is small, but for large values of $\alpha$, this term tends to zero at all points, except at the point where $cos2\chi l t = 1$ and its immediate vicinity. 
\end{enumerate}

Thus, at all time $\frac{j}{k} T_{rev}$ where $j = 1, 2,.., k-1$, the function shows a sudden peak. The matter of primary interest here being, these so called "signatures" are only noticed for the corresponding $k$ value. We see in the following sections that the corresponding plots of $\langle {\hat{O}} \rangle$, $\langle{\hat{O}^2}\rangle$ show signatures for the full revival(k = 1), and the first fractional revival(k = 2) respectively. (where $\hat{O}$ is some observable.) We can see that for subsequent plots the value of $k$ determines which fractional revival\cite{Wang2024WavePacket} can be seen in the plot, and the value of $k$ is determined by which moment of observable we wish to plot.\par In the succeeding sections, the expressions for $\langle{x(t)}\rangle$, $\langle{x^2(t)}\rangle$, $\langle{p(t)}\rangle$ and $\langle{p^2(t)}\rangle$ are calculated, and their plots are also provided. At the end of this chapter we shall look into other means of constructing the plots of moments of observables, which can be written as a function of the raising and lowering operator, using the \verb|qutip| module in Python. We shall show that constructing such plots require us to do as less as specifying the operators and setting up a suitable algorithm to plot the average values of these observables as a function of time. 
\par While one might argue that, the usage of such programming languages or numerical methods rob us off the opportunity to view the system in an analytic manner, the expressions for higher moments of observables can only be obtained analytically by performing redundant and tiresomely complex calculations, the end result of which is still its usage to construct the plot, which can be done almost effortlessly by using these algorithmic and numerical methods.

\subsection{Calculating $\langle{x(t)}\rangle$ and $\langle{p_x(t)}\rangle$}
  The signatures of the total revival can be seen in the plots of $\langle{x}\rangle$ and $\langle{p}\rangle$ . For our current discussions, we shall set $\hbar,m$ and $\omega$ to be $1$, and so from equations 1.2.4 and 1.2.5 ,$x = \frac{a + a^\dagger}{\sqrt{2}}$  and $p_x = \frac{a - a^\dagger}{i \sqrt{2}}$ , and using the general form , by setting k = 0; l = 1, we get
  
  \begin{equation}
  \langle{a}\rangle = \alpha \quad e^{-|\alpha|^2 (1 - cos2\chi t)}  e^{-i|\alpha|^2 sin2\chi t}
  \end{equation}
  and $\langle{a^{\dagger}}\rangle$ can be obtained by computing the complex conjugate of $\langle{a}\rangle$.
  \begin{equation}
      \langle{a^{\dagger}}\rangle = \alpha^* \quad e^{-|\alpha|^2(1-cos2\chi t)} e^{i |\alpha|^2 sin2\chi t}
  \end{equation}
  
  $\langle{x}\rangle$ is thus calculated to be:
  
  \begin{equation}
      \langle{x(t)}\rangle = \frac{1}{\sqrt{2}} \bra{\alpha(t)} (a + a^{\dagger}) \ket{\alpha(t)}
  \end{equation}
  
  \begin{equation}
      \langle{x(t)}\rangle = \frac{1}{\sqrt{2}} \bra{\alpha(t)}a\ket{\alpha(t)} + \bra{\alpha(t)}a^{\dagger}\ket{\alpha(t)}
  \end{equation}
  
  \begin{equation}
      \langle{x(t)}\rangle = \frac{1}{\sqrt{2}} (\alpha \quad e^{-|\alpha|^2 (1-cos2\chi t)} e^{-i |\alpha|^2 sin2\chi t}) + (\alpha^* \quad e^{-|\alpha|^2 (1-cos2\chi t)} e^{i|\alpha|^2 sin2\chi t})
  \end{equation}
  
\begin{equation}
    \langle{x(t)}\rangle = \frac{1}{\sqrt{2}} (e^{-|\alpha|^2 (1-cos2\chi t)}) (\alpha \quad e^{-i|\alpha|^2 sin2\chi t} + \alpha^* \quad e^{i|\alpha|^2 sin2\chi t})
\end{equation}
Let's assume the term $\alpha e^{-i|\alpha|^2 (sin2\chi t)}$ to be some complex number, say $\mathbb{Z}$, then it is evident that the complex conjugate of this value, that is, ${\mathbb{Z}}^*$ is nothing but, $\alpha^* e^{i|\alpha|^2 sin2\chi t}$. Hence the above equation can be written as:
\begin{equation}
    \langle{x(t)}\rangle = \frac{1}{\sqrt{2}} (e^{-|\alpha|^2(1-cos2\chi t)}) \{ \mathbb{Z} + \mathbb{Z}^* \} = \frac{1}{\sqrt{2}} (e^{-|\alpha|^2(1-cos2\chi t)}) (2\Re{\mathbb{Z}})
\end{equation}
Now the brunt of the problem lies in evaluating the real part of $\mathbb{Z}$.

\begin{equation}
    \Re{\mathbb{Z}} = \Re{\alpha \quad e^{-i|\alpha|^2 sin2\chi t}} 
\end{equation}
Recalling from equation (2.0.5) that $\alpha = \frac{p_1 + iq_1}{\sqrt{2}}$ , and expanding the latter term using the Euler's form, we get:
\begin{equation}
    \Re{\mathbb{Z}} = \Re{\left( \frac{p_1 + iq_1}{\sqrt{2}}\right) (cos(|\alpha|^2 sin2\chi t)) - i sin(|\alpha|^2 sin2\chi t)}
\end{equation}
Expanding this further:
\begin{equation}
    \Re{\mathbb{Z}} = \Re{\frac{1}{\sqrt{2}} \left(p_1cos(|\alpha|^2 sin2\chi t) -ip_1sin(|\alpha|^2 sin2\chi t)  + iq_1cos(|\alpha|^2 sin2\chi t)  +q_1sin(|\alpha|^2 sin2\chi t)  \right)}
\end{equation}

\begin{equation}
    \Re{\mathbb{Z}} = \frac{1}{\sqrt{2}} (p_1cos(|\alpha|^2 sin2\chi t) + q_1sin(|\alpha|^2 sin2\chi t))
\end{equation}
Substituting the above value in equation 3.1.7 we get:
  \begin{equation}
  \langle{x(t)}\rangle = e^{-|\alpha|^2 (1 - cos2\chi t)} [p_1cos(|\alpha|^2 sin2\chi t) + q_1sin(|\alpha|^2sin2\chi t)]
  \end{equation}
  
Similarly, we can get the value $\langle{p_x(t)}\rangle$ is:
  \begin{equation}
  \langle{p_x(t)}\rangle = e^{-|\alpha|^2 (1-cos2\chi t)} [-p_1sin(|\alpha|^2 sin2\chi t) + q_1cos(|\alpha|^2sin2\chi t)]
  \end{equation}

  from which we readily see that $\langle{x(t=0)}\rangle = p_1$ and $\langle{p_x(t=0)}\rangle = q_1$ .
  Due to the exponential term, the expectation value of x remains near $p_1$ for most of the time, jumping suddenly to rapid variations at the instants of full revival.(the same holds for $p_x$)
  This phenomenon is seen very prominently when we set $p_1$ and $q_1$ to large values. The Hamiltonian has a symmetry as we can see from the values of $\langle{x}\rangle$ and $\left\langle{p_x}\right\rangle$, and by setting $p_1 = q_1$, we can without loss of generality, proceed with our study.
  Here are the plots for $\langle{x}\rangle$ and $\langle{p_x}\rangle$ for three different values of $p_1,q_1$ .
  
  \begin{figure}[H]
    \centering
    \includegraphics[width=15cm]{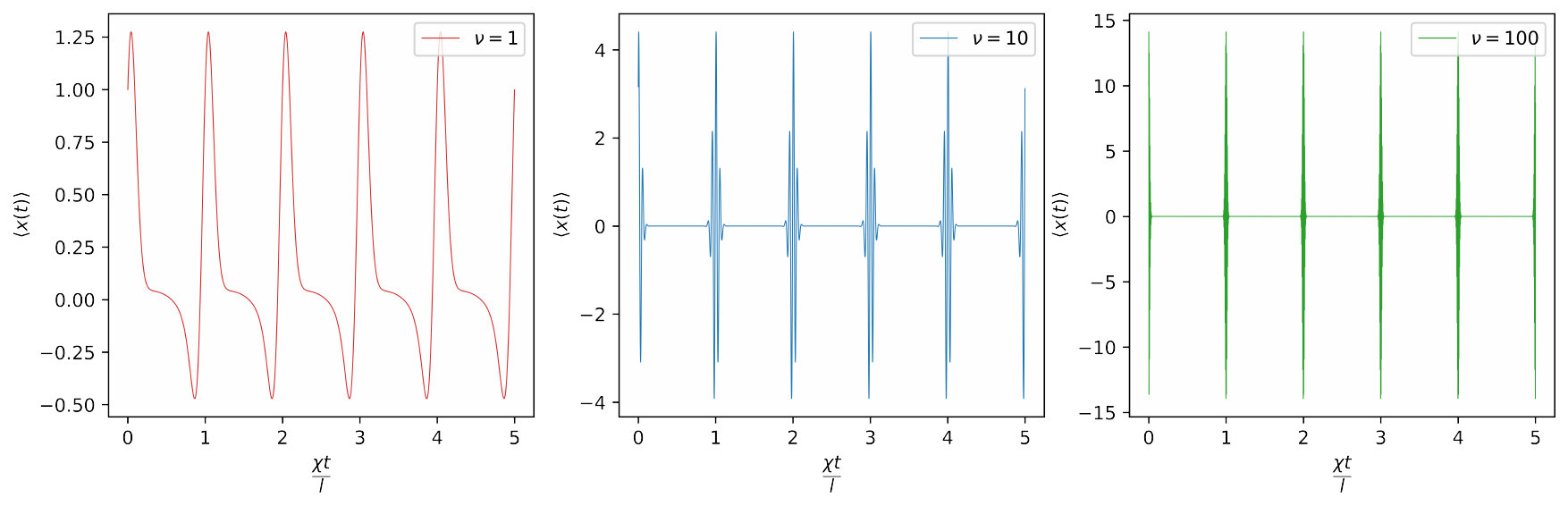}
    \label{fig:only_x}
    \caption{Plots of $\langle{x(t)}\rangle$ vs $\frac{\chi t}{\pi}$, for i) $\nu = 1$, ii) $\nu = 10$, iii)$\nu = 100$}
\end{figure}

  \begin{figure}[H]
    \centering
    \includegraphics[width=15cm]{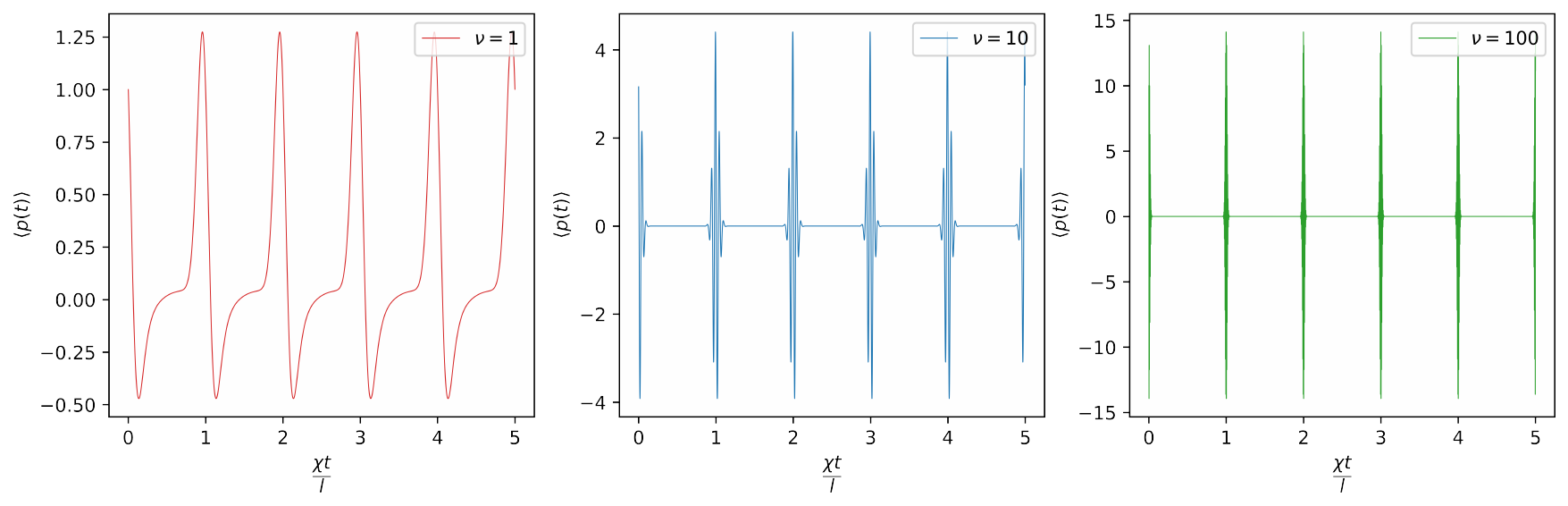}
    \label{fig:only_p(1)}
    \caption{Plots of $\langle{p(t)}\rangle$ vs $\frac{\chi t}{\pi}$, for i) $\nu = 1$, ii) $\nu = 10$, iii)$\nu = 100$}
\end{figure}

\subsubsection{"Phase plots"}  
 While it is evident here, that constructing phase plots for quantum systems is hitherto unprecedented, as in the quantum regime, both position and momentum are only seen as operators, we venture to do exactly, because we have considered coherent states here, which behave under certain conditions like a "classical system.
  From the static images of the "phase plots"\cite{morin_waves_chapter6_dispersion}, we cannot see the path traced out by the representative point as time evolves. The ]links for the animated plots are provided in Appendix B. \newline
  
  The representative point remains inactive and at the origin at all times except the time intervals when it approaches the revival time $t=T_{rev}$, at which point it rapidly scales the remaining curve and then comes back again to the origin.
  
  \begin{figure}[H]
    \centering
    \includegraphics[width=15cm]{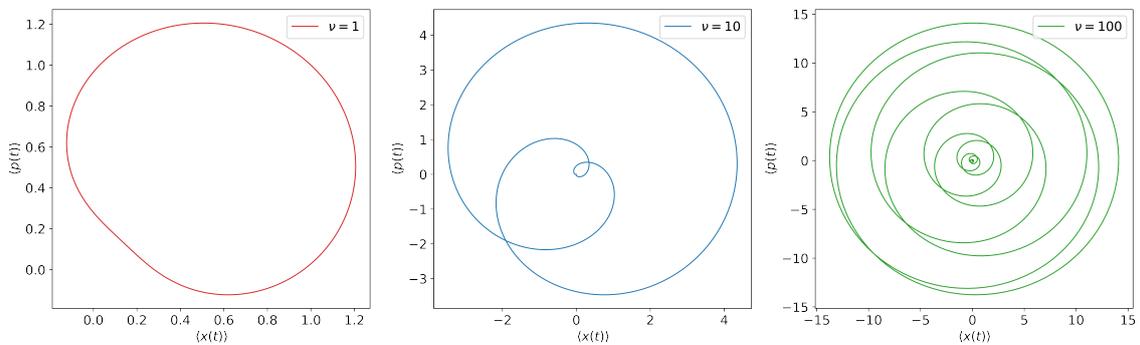}
    \label{fig:phase_plot}
    \caption{Phase Plots of $\langle{p(t)}\rangle$ vs $\langle{x(t)}\rangle$, for i) $\nu = 1$, ii) $\nu = 10$, iii)$\nu = 100$}
\end{figure}
  
  \subsubsection{Second Moment Of $\left\langle{x}\right\rangle$ AND $\langle{p_x}\rangle$}
From the equation 2.0.8, by setting $k=2$ and $l=0$ we can obtain 
\begin{equation}
\langle{a^2(t)}\rangle = \alpha^2 e^{-|\alpha|^2 (1 - cos4\chi t)} exp[-i2\chi t - i|\alpha|^2sin4\chi t]
\end{equation}
and its complex conjugate:
\begin{equation}
\langle{a^{\dagger 2}(t)}\rangle = \alpha^{* 2} e^{-|\alpha|^2(1-cos4\chi t)} exp[i2\chi t + i|\alpha|^2 sin4\chi t] 
\end{equation}
and finally, the expression for $\langle{a^{\dagger}a}\rangle$ is obtained by setting $k = 1$ and $l=0$:
\begin{equation}
    \langle{a^{\dagger}a(t)}\rangle = \alpha^0 |\alpha|^2 e^{-|\alpha|^2 (1-cos2(0)\chi t)} exp[-i \chi ((0)(-1) + 2 (1)(0) )t - i |\alpha|^2 2 (0)\chi t] = |\alpha|^2 
\end{equation}

 from which we get $\langle{x^2(t)}\rangle$ and $\langle{p^2_x(t)}\rangle$ by almost the same method that we did to find $\langle{x}\rangle$ and $\langle{p_x}\rangle$.
 We begin by stating the following:
 \begin{equation}
     \langle{x^2(t)}\rangle = \frac{1}{2} \bra{\alpha(t)}{(a + a^{\dagger})^2}\ket{\alpha(t)}
 \end{equation}
 \begin{equation}
     \langle{x^2(t)}\rangle = \frac{1}{2}[ \langle{a^2(t)}\rangle + \langle{a^{\dagger 2}(t)}\rangle + 2 \langle{a^{\dagger} a (t)}\rangle + 1] 
 \end{equation}
 
 \begin{multline}
     \langle{x^2(t)}\rangle = \frac{1}{2} \{ e^{-|\alpha|^2 (1-cos4\chi t)} (\alpha^2 exp(-i2\chi t -i|\alpha|^2 sin4\chi t)\\ + \alpha^{* 2} exp(i2 \chi t + i |\alpha|^2 sin4\chi t) + 2|\alpha|^2 + 1  ) \}
 \end{multline}
 
We again see that the two complex terms are conjugate to each other and can be re-written as follows:

\begin{equation}
    \langle{x^2 (t)}\rangle = \frac{1}{2} \{ e^{-|\alpha|^2 (1-cos4\chi t)}[1 + 2|\alpha|^2 + 2 \Re{\alpha^2 exp(-(i2\chi t + |\alpha|^2 sin4\chi t))}] \}
\end{equation}
As before, we shall now calculate the real part of $\alpha^2 exp(-i(2\chi t + |\alpha|^2 sin4\chi t))$:
\begin{equation}
    \alpha^2 exp(-i(2\chi t + |\alpha|^2 sin4\chi t)) = (p_1 + iq_1)^2 [cos(2\chi t + |\alpha|^2 sin4\chi t) -isin(2\chi t + |\alpha|^2 sin4\chi t) ]
\end{equation}

\begin{multline}
    \alpha^2 exp(-i(2\chi t + |\alpha|^2 sin4\chi t)) = (p_1^2 - q_1^2 +i2p_1q_1) [cos(2\chi t + |\alpha|^2 sin4\chi t) \\-isin(2\chi t + |\alpha|^2 sin4\chi t) ]
\end{multline}
And so, we can write the real part of the above expression as:
\begin{multline}
    \Re{\alpha^2 exp(-i(2\chi t + |\alpha|^2 sin4\chi t))} = (p_1^2 - q_1^2) cos(2\chi t + |\alpha|^2 sin4\chi t)\\ + 2p_1q_1 sin(2\chi t + |\alpha|^2 sin4\chi t)
\end{multline}
Substituting this in equation (2.2.7):
\begin{multline}
2\langle{x^2(t)}\rangle = 1+p_1^2 +q_1^2 + e^{-|\alpha|^2 (1-cos4\chi t) }[(p_1^2 - q^2_1) cos(2\chi t + |\alpha|^2 sin4\chi t) \\ + 2p_1 q_1 sin(2\chi t + |\alpha|^2 sin4\chi t)]
\end{multline}
Similarly, we can obtain the expression for $\langle{p^2(t)}\rangle$ :

\begin{multline}
2\langle{p_x^2(t)}\rangle = 1+p_1^2 +q_1^2 -e^{-|\alpha|^2 (1-cos4\chi t) }[(p_1^2 - q^2_1) cos(2\chi t + |\alpha|^2 sin4\chi t) \\ + 2p_1 q_1 sin(2\chi t + |\alpha|^2 sin4\chi t)]
\end{multline}

The plots for $i) p_1,q_1 = 1$  ;$ii) p_1,q_1 = \sqrt{10}$ ;$iii) p_1,q_1 = 10$ are given below for  $\left\langle{x^2}\right\rangle$ and $\left\langle{p^2_x}\right\rangle$, wherein we notice rapid variations at and near the time $t = \frac{T_{rev}}{2}$, indicative of the first fractional revival.

 \begin{figure}[H]
    \centering
    \includegraphics[width=15cm]{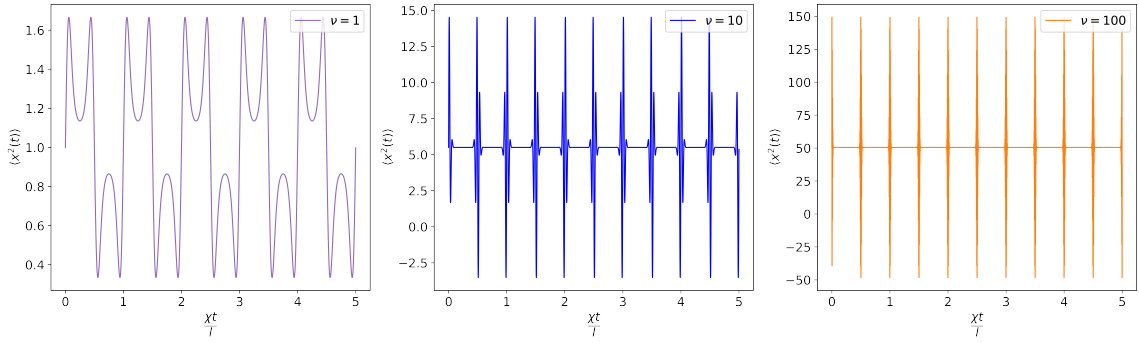}
    \label{fig:only_x2}
    \caption{Plots of $\langle{x^2(t)}\rangle$ vs $\frac{\chi t}{\pi}$, for i) $\nu = 1$, ii) $\nu = 10$, iii)$\nu = 100$}
\end{figure}

 \begin{figure}[H]
    \centering
    \includegraphics[width=15cm]{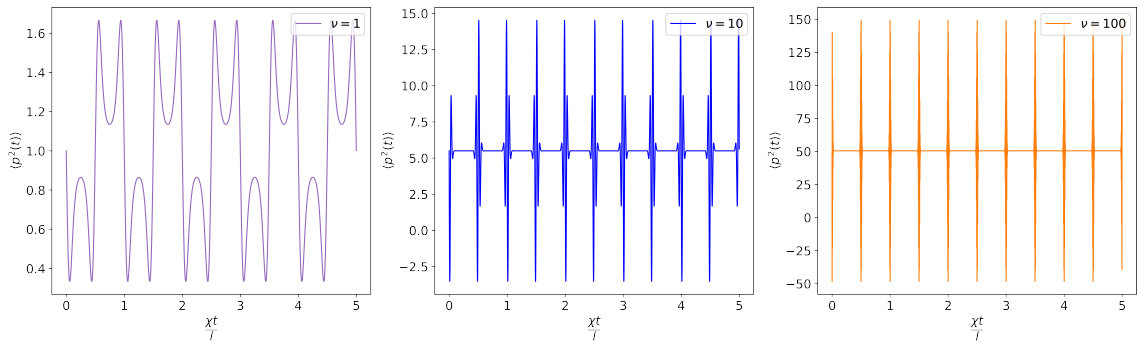}
    \label{fig:only_p2}
    \caption{Plots of $\langle{p^2(t)}\rangle$ vs $\frac{\chi t}{\pi}$, for i) $\nu = 1$, ii) $\nu = 10$, iii)$\nu = 100$}
\end{figure}

With the quantities thus calculated so far, we can venture out to obtain the uncertainty product $\Delta{x} \Delta{p}$ .
We can plot $\Delta{x} \Delta{p} $ vs  time plots and $\Delta{p}$ vs $\Delta{x}$ plots(the so-called "phase-plots") for the same $p_1$ and $q_1$ values.

\begin{figure}[H]
    \centering
    \includegraphics[width=15cm]{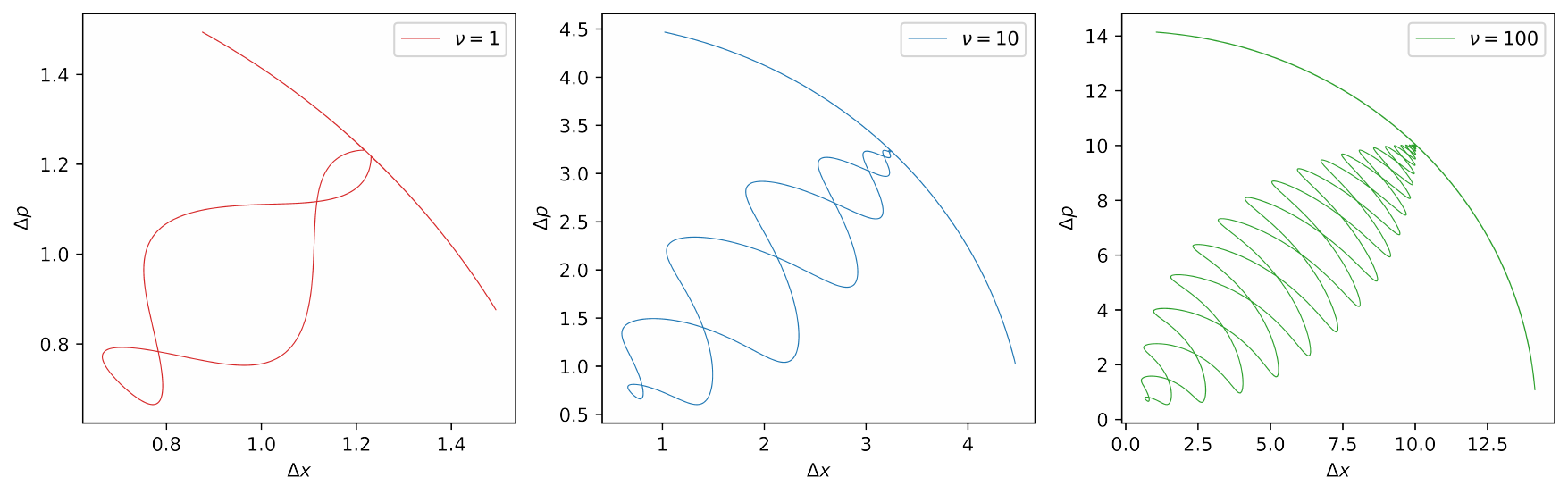}
    \label{fig:phase}
    \caption{Plots of $\Delta{p}$ vs $\Delta{x}$, for i) $\nu = 1$, ii) $\nu = 10$, iii)$\nu = 100$}
\end{figure}

\subsection{Using the $\text{Qutip}$ module to construct the plots of observables}

\par While working on computing the expressions for these observables, it was tempting for us to envision an algorithm which would only require for us to enter the Hamiltonian and the operator as a function of the raising and lowering operators. While searching for such algorithms, we chanced upon the \verb|qutip| module(programmed in Python) which allows us to write programs which require the Hamiltonian, and the number of Fock states, and certain specifics of the initial coherent state.

\par \textit{To use} \verb|qutip|, \textit{one only needs to have some knowledge of the basic workings of Python, and the} \verb|numpy| \textit{module.} 

\par The entire program will not be provided here, but certain functions and methods used, shall be highlighted. The states, operators which have been initialised in \verb|qutip| have an added functionality, as they are defined as a Quantum Object (a method), or \verb|qutip.Qobj()|\cite{qutip_basics_guide}, which take in only row, column matrices (which are identified as "bra" and "ket" respectively) and square matrices(identifies it as "operator"). The method identifies the type and in the case of an operator, it returns a boolean output, which tells us whether the square matrix is hermitian or not.

The module has predefined functions for initializing Fock states, coherent states, pauli matrices, raising and lowering operators and much more\cite{qutip_api_functions}. For eg., let us initialize a raising operator, with a basis of 10 energy eigen states.
The required method is:
\newline
\verb|qutip.create(N = Number of Fock states in Hilbert Space)|

\begin{tcolorbox}
\begin{lstlisting}[language = Python]
from qutip import *
a = create(10)
print(a)
\end{lstlisting}
\end{tcolorbox}

The output of the above snippet is:

\begin{tcolorbox}

Quantum object: dims = [[10], [10]], shape = (10, 10), type = oper, isherm = False\begin{equation*}\left(\begin{array}{*{11}c}0.0 & 0.0 & 0.0 & 0.0 & 0.0 & 0.0 & 0.0 & 0.0 & 0.0 & 0.0\\1.0 & 0.0 & 0.0 & 0.0 & 0.0 & 0.0 & 0.0 & 0.0 & 0.0 & 0.0\\0.0 & 1.414 & 0.0 & 0.0 & 0.0 & 0.0 & 0.0 & 0.0 & 0.0 & 0.0\\0.0 & 0.0 & 1.732 & 0.0 & 0.0 & 0.0 & 0.0 & 0.0 & 0.0 & 0.0\\0.0 & 0.0 & 0.0 & 2.0 & 0.0 & 0.0 & 0.0 & 0.0 & 0.0 & 0.0\\0.0 & 0.0 & 0.0 & 0.0 & 2.236 & 0.0 & 0.0 & 0.0 & 0.0 & 0.0\\0.0 & 0.0 & 0.0 & 0.0 & 0.0 & 2.449 & 0.0 & 0.0 & 0.0 & 0.0\\0.0 & 0.0 & 0.0 & 0.0 & 0.0 & 0.0 & 2.646 & 0.0 & 0.0 & 0.0\\0.0 & 0.0 & 0.0 & 0.0 & 0.0 & 0.0 & 0.0 & 2.828 & 0.0 & 0.0\\0.0 & 0.0 & 0.0 & 0.0 & 0.0 & 0.0 & 0.0 & 0.0 & 3.0 & 0.0\\\end{array}\right)\end{equation*}
\end{tcolorbox}

The \verb|Qobj| denotes that the raising operator is a non-Hermitian matrix, with 10 rows and 10 columns.

A coherent state, whose eigen value when operated by a lowering operator is $\alpha = 1+1i$, and whose basis are 10 energy eigen states, is initialised with the following method.
\vskip 2em
\verb|qutip.coherent(N=No.of Fock States, alpha=eigen value of lowering operator)|
\newline
(Note: These two are the essential arguments, and there are other arguments\cite{qutip_api_functions} which can be given by the user, but we shall not discuss those here. This applies for all the subsequent functions discussed in this section.)

\begin{tcolorbox}
\begin{lstlisting}[language = Python]
c = coherent(10, 1 + 1j)
print(c)
\end{lstlisting}
\end{tcolorbox}

\begin{tcolorbox}
\textbf{OUTPUT}
Quantum object: dims = [[10], [1]], shape = (10, 1), type = ket\begin{equation*}\left(\begin{array}{*{11}c}0.368\\(0.368+0.368j)\\0.520j\\(-0.300+0.300j)\\-0.300\\(-0.134-0.134j)\\-0.110j\\(0.042-0.042j)\\0.028\\(0.012+0.012j)\\\end{array}\right)\end{equation*}
\end{tcolorbox}

The matrix gives us the co-efficients of the energy eigen states which constitiute the coherent state.

\textbf{Although one might argue that, the very definition of a coherent state is that it is a superposition of infinite energy eigenstates, whereas N here, is some finite value. In the case of numerical evaluation, it might suffice us to confine ourselves to some finite value, owing to computational limitations, and also to the fact that with such a finite number of Fock states one can simulate a coherent state with a satisfiable degree of accuracy.}

The next step would be to initialise the Hamiltonian $H = a^{\dagger 2} a^2 \chi$. By following the conventional notation, we shall use \verb|a| to represent the lowering operator, and its complex conjugate \verb|a.dag()| (where \verb|dag()| method gives us the complex conjugate of a \verb|Qobj|) to represent the raising operator. We shall set $\chi = \frac{10}{\pi}$.

\begin{tcolorbox}
\begin{lstlisting}[language = Python]
import numpy as np

a = destroy(200)
chi = 10 / np.pi

H = chi * (a.dag()**2 * a**2)
print(H)
\end{lstlisting}
\end{tcolorbox}
The Hamiltonian in matrix form is given as output.

\small{Quantum object: dims = [[200], [200]], shape = (200, 200), type = oper, isherm = True\begin{equation*}\left(\begin{array}{*{11}c}0.0 & 0.0 & 0.0 & 0.0 & \cdots & 0.0 & 0.0 & 0.0 & 0.0\\0.0 & 0.0 & 0.0 & 0.0 & \cdots & 0.0 & 0.0 & 0.0 & 0.0\\0.0 & 0.0 & 6.366 & 0.0 & \cdots & 0.0 & 0.0 & 0.0 & 0.0\\0.0 & 0.0 & 0.0 & 19.099 & \cdots & 0.0 & 0.0 & 0.0 & 0.0\\0.0 & 0.0 & 0.0 & 0.0 &  \cdots & 0.0 & 0.0 & 0.0 & 0.0\\\vdots & \vdots & \vdots & \vdots & \ddots & \vdots & \vdots & \vdots & \vdots\\0.0 & 0.0 & 0.0 & 0.0 & \cdots & 0.0 & 0.0 & 0.0 & 0.0\\0.0 & 0.0 & 0.0 & 0.0 & \cdots & 1.217\times10^{+05} & 0.0 & 0.0 & 0.0\\ 0.0 & 0.0 & 0.0 & 0.0 & \cdots & 0.0 & 1.229\times10^{+05} & 0.0 & 0.0\\0.0 & 0.0 & 0.0 & 0.0 & \cdots & 0.0 & 0.0 & 1.242\times10^{+05} & 0.0\\0.0 & 0.0 & 0.0 & 0.0 & \cdots & 0.0 & 0.0 & 0.0 & 1.254\times10^{+05}\\\end{array}\right)\end{equation*}}

The \verb|qutip| module has an inbuilt function \verb|qutip.mesolve()| which allows us to \textbf{\textit{numerically}} obtain the expectation value of some observable as a function of time. The function requires the following necessary values as arguments.

\begin{enumerate}
    \item The Hamiltonian
    \item The State vector(Either "bra" or "ket" state)
    \item An array, which begins and ends with the initial and final time (as specified by the user), divided into $n$ elements (where $n$ is the step-size.)
\end{enumerate}

Let us initialise a coherent state containing 200 energy eigen states, and for which $\alpha = 10$, and also create a numpy array called \verb|steps| beginning at $t=0$, and ending at $t=5$ seperated into 10000 steps, using the \verb|numpy.linspace()| function. We shall use the \verb|qutip.mesolve()| function, and store the result in a variable called \verb|result|.

\begin{tcolorbox}
\begin{lstlisting}[language = Python]
co = coherent(200, 10)
steps = np.linspace(0, 5, 10000) 

result = mesolve(H, co, steps)
\end{lstlisting}
\end{tcolorbox}

The \verb|result| variable contains a 2D numpy array whose rows (1000 in number) contain the time evolved state vectors (each of which, as specified previously consists of 200 eigen states.)

One can view the time evolved state vector by using the following command \verb|result.states[n]| where n should be less than or equal to the no.of steps (in our case, 10000). Let's look at the $5000^{th}$ state vector in the \verb|result| array, or in other words, the state vector at time $t = \pi$:

\begin{tcolorbox}
\begin{lstlisting}[language = Python]
print(result.states[5000])

\end{lstlisting}
\end{tcolorbox}
\textbf{OUTPUT}
\begin{tcolorbox}
Quantum object: dims = [[200], [1]], shape = (200, 1), type = ket\begin{equation*}\left(\begin{array}{*{11}c}0.135\\0.271\\(0.149-0.353j)\\(-0.412+0.160j)\\(0.327-0.298j)\\\vdots\\0.0\\0.0\\0.0\\0.0\\0.0\\\end{array}\right)\end{equation*}
\end{tcolorbox}

Now, that the numerical solution has been obtained, the next step is to let an operator act upon these time-evolved states. Let us begin with the position operator $\hat{x} = \frac{1}{\sqrt{2}} (a + a^{\dagger})$.We will initialize a variable \verb|x|, as this position operator. The functions \verb|qutip.expect()| and \verb|qutip.variance()| will be used by us, to find the expectation value and the variance of the operator, and these values shall be stored in the variables \verb|x_op| and \verb|x_var| respectively. Both these functions require the following inputs as arguments.

\begin{enumerate}
    \item The operator
    \item The 2D array which can be called by \verb|result.states|
\end{enumerate}

\begin{tcolorbox}
\begin{lstlisting}[language = Python]
x = np.sqrt(0.5) * (a + a.dag())

x_op = expect(x, result.states)
x_var = variance(x, result.states)
\end{lstlisting}
\end{tcolorbox}
The rest of the program, only requires us to plot these two arrays \verb|x_op| and \verb|x_var|, which can be done using any plotting module available in Python, like \verb|matplotlib|, \verb|plotly|, \verb|seaborn|, etc.,

The plot of $\langle{x}\rangle$ vs time, and $(\triangle{x})^2$ vs time are given together below.

\begin{figure}[H]
    \centering
    \includegraphics[width=7cm]{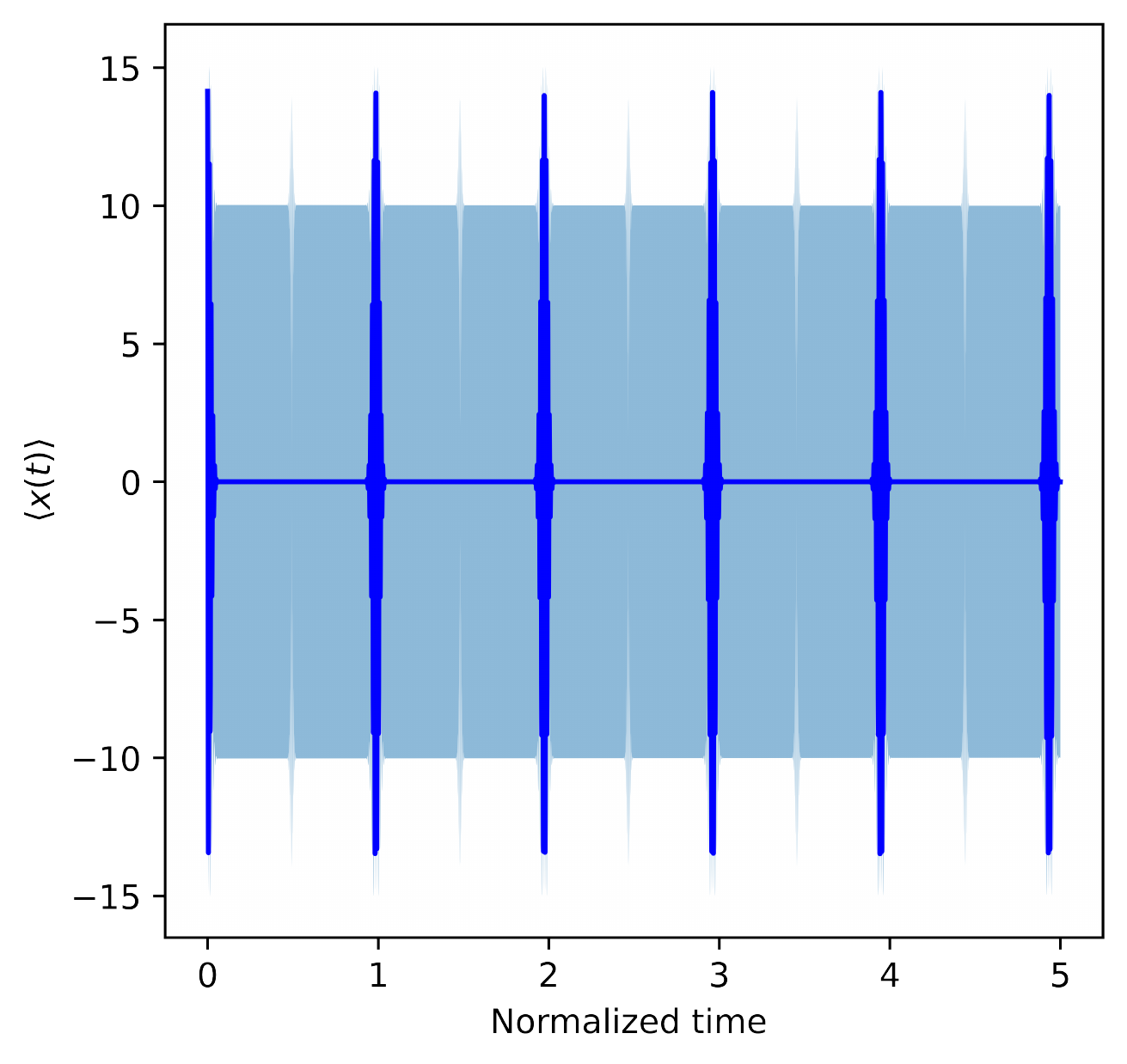}
    \label{fig:only_lx}
    \caption{Plots of $\langle{x(t)}\rangle$ vs time and $(\triangle{x})^2$ vs time for $\alpha = 10$}
\end{figure}

The plot obtained thus, agrees well with that constructed using the expression obtained through analytic means. Let's try to do the same for the operator $\langle{x^3(t)}\rangle$, which in theory, must show us the signature for the second fractional revival.

\begin{figure}[H]
    \centering
    \includegraphics[width=7cm]{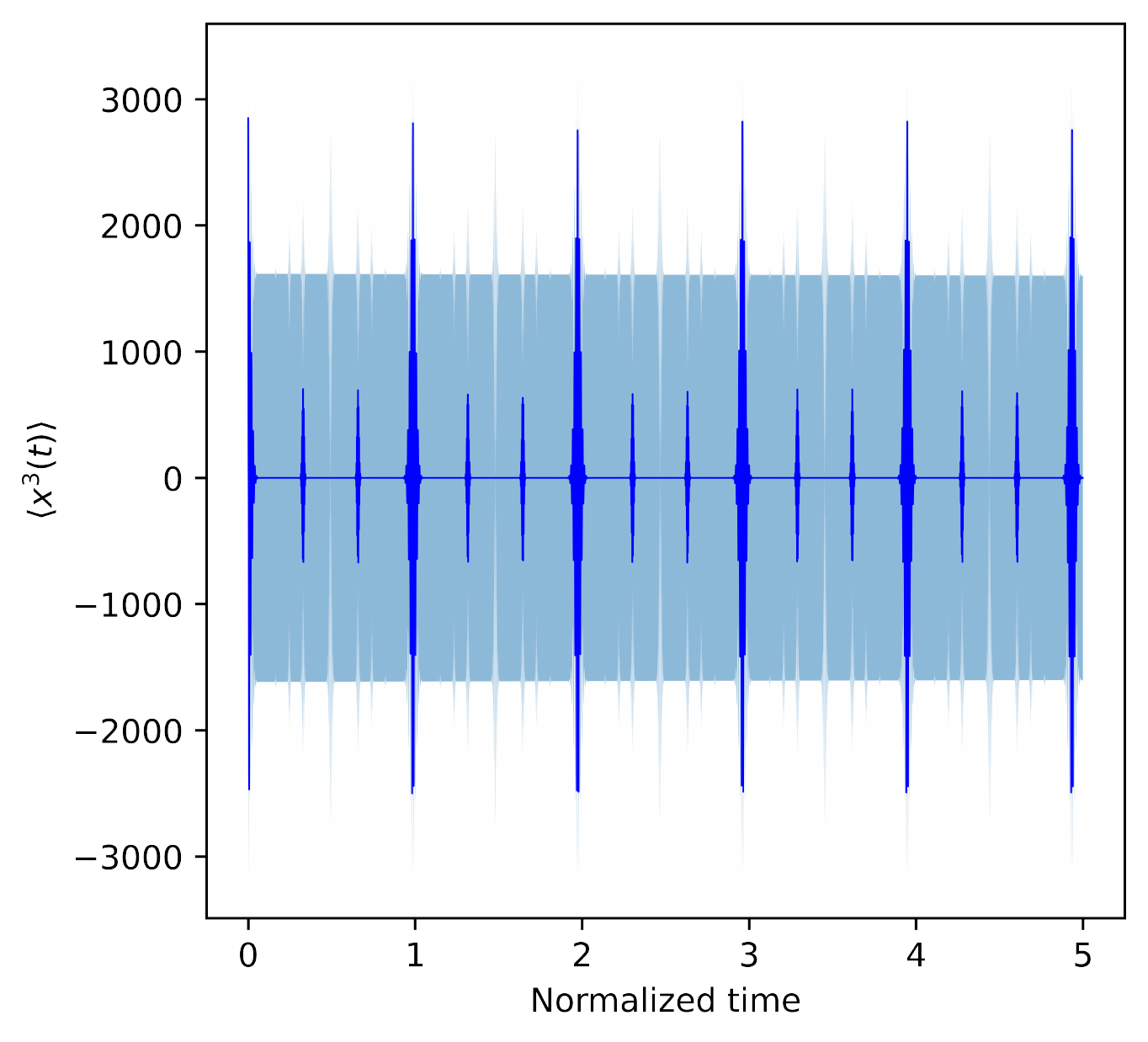}
    \label{fig:only_lx}
    \caption{Plots of $\langle{x^3(t)}\rangle$ vs time and $(\triangle{x^3})^2$ vs time for $\alpha = 10$}
\end{figure}

\subsection{Angular Momentum}

Extending the argument to three dimensions, we can write the three dimensional non-linear Hamiltonian as:
\begin{equation}
H =  (a^{\dagger 2} a^2+b^{\dagger 2} b^2+c^{\dagger 2} c^2) \hbar \chi
\end{equation}
where $a$ and $a^{\dagger}$, $b$ and $b^{\dagger}$, $c$ and $c^{\dagger}$ are the annihilation and creation operators along the three orthogonal axes respectively. Also, it is convenient here to state that $\alpha,\beta,\gamma$ are the eigen values of the lowering operators $a,b,c$ , additionally: 

\begin{equation}
\alpha = \frac{p_1 + iq_1}{\sqrt{2}} = r_1 e^{i \theta_1}
\end{equation}

\begin{equation}
\beta = \frac{p_2 + i q_2}{\sqrt{2}} = r_2 e^{i \theta_2} 
\end{equation}

\begin{equation}
\gamma = \frac{p_3 + i q_3}{\sqrt{2}} = r_3 e^{i \theta_3} 
\end{equation}

The three dimensional time-dependent coherent state can be written as:
\begin{equation}
\ket{\psi_{\alpha}} \otimes \ket{\psi_{\beta}} \otimes \ket{\psi_{\gamma}} =\ket{\psi_{\alpha.\beta,\gamma}} = e^{-(|\alpha|^2 + |\beta|^2 + |\gamma|^2) /2} \sum^{\infty}_{n,l,m=0} \frac{\alpha^n \beta^l \gamma^m}{\sqrt{n!l!m!}} e^{-i \chi H t} \ket{n,l,m} 
\end{equation}

 We know that $\overrightarrow{L} = \overrightarrow{r} \times \overrightarrow{p}$ , and so $L_x = yp_z - zp_y$. Substituting the values
 $y = \frac{b+b^{\dagger}}{\sqrt{2}}$, $z = \frac{c+c^{\dagger}}{\sqrt{2}}$, $p_y = \frac{b-b^{\dagger}}{i\sqrt{2}}$, $p_z = \frac{c-c^{\dagger}}{i\sqrt{2}}$ we get:
 \begin{equation}
 L_x = \frac{1}{2i} (b^{\dagger}c - c^{\dagger}b)
 \end{equation}
and similarly,
\begin{equation}
L_y = \frac{1}{2i} (c^{\dagger}a - a^{\dagger}c)
\end{equation}
\begin{equation}
L_z = \frac{1}{2i} (a^{\dagger}b - b^{\dagger}a)
\end{equation}
For our current discussion, we shall restrict ourselves to $L_x$. It can be computed from (1). We set the values $\beta = p_2 + iq_2$ and $\gamma = p_3 + iq_3$:
\subsubsection{First Moment of $L_x$}
The expression $\langle{L_x(t)}\rangle$ can be evaluated as follows:

\begin{equation}
    \langle{L_x}\rangle = \frac{-i}{2} [\langle{b^{\dagger}}\rangle \langle{c}\rangle - \langle{c^\dagger}\rangle \langle{b}\rangle]
\end{equation}

The values $\langle{b}\rangle$, $\langle{b^\dagger}\rangle$, $\langle{c}\rangle$ and $\langle{c^\dagger}\rangle$ are the following:

\begin{equation}
  \langle{b}\rangle = \beta e^{-|\beta|^2 (1 - cos2\chi t)}  e^{-i|\beta|^2 sin2\chi t}
  \end{equation}
  
\begin{equation}
  \langle{b^\dagger}\rangle = \beta^* e^{-|\beta|^2 (1 - cos2\chi t)}  e^{i|\beta|^2 sin2\chi t}
  \end{equation}
  
\begin{equation}
  \langle{c}\rangle = \gamma e^{-|\gamma|^2 (1 - cos2\chi t)}  e^{-i|\gamma|^2 sin2\chi t}
  \end{equation}
  
\begin{equation}
  \langle{c^\dagger}\rangle = \gamma^* e^{-|\gamma|^2 (1 - cos2\chi t)}  e^{i|\gamma|^2 sin2\chi t}
  \end{equation}
as can be easily obtained from the general form. 

We can thus construct the expression for $\langle{L_x}\rangle$ from the above four expressions:

\begin{multline}
    \langle{L_x}\rangle = \frac{-i}{2}  [  \beta^* e^{-|\beta|^2 (1 - cos2\chi t)}  e^{i|\beta|^2 sin2\chi t} \times \gamma e^{-|\gamma|^2 (1 - cos2\chi t)}  e^{-i|\gamma|^2 sin2\chi t} \\ -  \gamma^* e^{-|\gamma|^2 (1 - cos2\chi t)}  e^{i|\gamma|^2 sin2\chi t} \times  \beta e^{-|\beta|^2 (1 - cos2\chi t)}  e^{-i|\beta|^2 sin2\chi t}]
\end{multline}

\begin{multline}
    \langle{L_x}\rangle = \frac{-i}{2}  [ (e^{(|\beta|^2 + |\gamma|^2)(1-cos2\chi t)})\{ \beta^*  e^{i|\beta|^2 sin2\chi t}  \gamma  e^{-i|\gamma|^2 sin2\chi t} -  \gamma^* e^{i|\gamma|^2 sin2\chi t}  \beta e^{-i|\beta|^2 sin2\chi t}\}]
\end{multline}

\begin{multline}
    \langle{L_x}\rangle = \frac{-i}{2}  [ (e^{(|\beta|^2 + |\gamma|^2)(1-cos2\chi t)})\{ \beta^* \gamma  e^{i(|\beta|^2 - |\gamma|^2) sin2\chi t}  - \beta \gamma^* e^{i(|\gamma|^2 - |\beta|^2) sin2\chi t} \}]
\end{multline}

\begin{equation}
    \langle{L_x}\rangle = \frac{-i}{2}  (e^{(|\beta|^2 + |\gamma|^2)(1-cos2\chi t)})(2i) \Im{\beta^* \gamma  e^{i(|\beta|^2 - |\gamma|^2) sin2\chi t}}
\end{equation}

We find that the calculation is similar to the previous observables, but only with a slight increase in complexity.
Henceforth, we shall indulge ourselves in computing the imaginary part of the expression $\beta^* \gamma e^{i(|\beta|^2 -|\gamma|^2) sin2\chi t}$. For ease of calculation let us set $x = (|\beta|^2 - |\gamma|^2) sin2\chi t$

\begin{equation}
\beta^* \gamma e^{ix} = \frac{1}{2}(p_2 - iq_2)(p_3+iq_3)(cos(x) + isin(x)) 
\end{equation}

\begin{multline}
    \beta^* \gamma e^{ix} = \frac{1}{2}\{p_2 q_3 cos(x) + i p_2 q_3 sin(x) + i p_2 q_3 cos(x) - p_2 q_3 sin(x)\\ - i q_2 p_3 cos(x) + q_2 p_3 sin(x) + q_2 q_3 sin(x) + i q_2 q_3 sin(x)\}
\end{multline}

But we are only interested in the imaginary part of the above expression.

\begin{equation}
    \Im{\beta^* \gamma e^{ix}} = \frac{1}{2} \{(p_2 q_3 + q_2 q_3)sin(x) + (p_2 q_3 - q_2 p_3) cos(x)\}
\end{equation}

Replacing $x$ by the original value, and substituting this in the expression for $\langle{L_x}\rangle$, we get:
\begin{multline}
    \langle{L_x}\rangle = \frac{1}{2}  (e^{(|\beta|^2 + |\gamma|^2)(1-cos2\chi t)}) \{ \frac{1}{2}(p_2 q_3 + q_2 q_3)sin((|\beta|^2 - |\gamma|^2) sin2\chi t) \\ + (p_2 q_3 - q_2 p_3) cos((|\beta|^2 - |\gamma|^2) sin2\chi t)\}
\end{multline}

We might as well replace $\beta$ by $\frac{p_2^2 + q_2^2}{2}$ and $\gamma$ by $\frac{p_3^2 + q_3^2}{2}$:
\begin{multline}
2\langle{L_x}\rangle = e^{-\frac{1}{2}[(p_2^2 + q^2_2)+(p_3^2 + q^2_3)] (1-cos2\chi t)}  [(p_2 q_3 + q_2 q_3)sin(((\frac{p_2^2 + q^2_2}{2})-(\frac{p_3^2 + q_3^2}{2})) sin 2\chi t) \\ + (p_2 q_3 - q_2 p_3)cos(((\frac{p_2^2 + q^2_2}{2}) - (\frac{p^2_3 + q^2_3}{2}))sin2\chi t)]
\end{multline}

The plots for $\langle{L_x}\rangle$ vs time, are given below.

\begin{figure}[H]
    \centering
    \includegraphics[width=15cm]{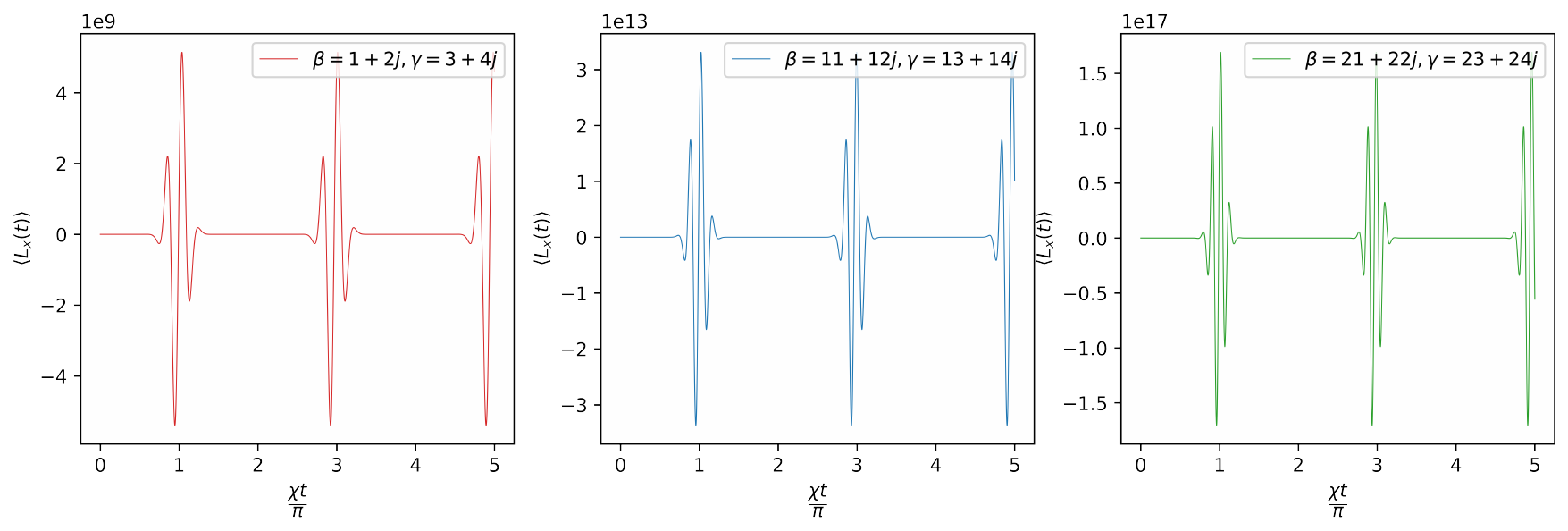}
    \label{fig:only_lx}
    \caption{Plots of $\langle{L_x(t)}\rangle$ vs $\frac{\chi t}{\pi}$, for i) $\beta = 1+2j, \gamma = 3+4i$, ii) $\beta = 11 + 12j, \gamma = 13 + 14j$, iii)$\beta = 21 + 22j, \gamma = 23 + 24j$.}
\end{figure}

$\langle{L_x}\rangle$ mimics the behaviour of $\langle{x}\rangle$ and $\langle{p_x}\rangle$, by only showing signatures of the full revival, by bursting into rapid variations at $t = T_{rev}$ .
Conversely, one can write $\langle{L_x}\rangle$ in terms of $\beta = r_2 e^{i\theta_2}$ and $\gamma = r_3 e^{i\theta_3}$ in the following way:
\begin{equation}
\langle{L_x}\rangle = r_2 r_3 e^{-((r_2^2 +r_3^2) (1 - cos2\chi t))} [sin((\theta_3 - \theta_2)(r_2^2 -r_3^2)sin2\chi t)]
\end{equation}
and we shall define $\langle{L_x^2}\rangle$ in the above given method. When the terms $p_2$ and $q_3$, $p_3$ and $q_2$ are equal to each other, the cosine part of the equation vanishes completely. And when $p_2 = q_2$ and $p_3 = q_3$, then the exponential part also reduces to unity, and the frequency terms of the sine and cosine terms become zero, thus, showing no variation throughout. 

Hence, when:
\begin{equation}
    p_2 = q_2, p_3 = q_3; \langle{L_x(t)}\rangle = 0
\end{equation}
\subsubsection{Second Moment of $L_x$}

The expression for $\langle{L_x^2}\rangle$ is given by:

\begin{equation}
    L_x^2 = \frac{-1}{4} (b^{\dagger}c - c^{\dagger}b)^2
\end{equation}

\begin{equation}
    L_x^2 = \frac{-1}{4} (b^{\dagger 2}c^2 - b^\dagger b cc^\dagger - c^\dagger c b b^\dagger + c^{\dagger 2}b^2)
\end{equation}

\begin{equation}
    L_x^2 = \frac{-1}{4} (b^{\dagger 2}c^2 + c^{\dagger 2}b^2 -b^{\dagger} b (c^{\dagger}c + 1) - c^\dagger c (b^\dagger b + 1))
\end{equation}
Finding the average value of the above expression when acted upon a coherent state, we get

\begin{equation}
    \langle L_x^2\rangle = \frac{-1}{4} (\langle b^{\dagger 2}\rangle \langle{c^2}\rangle + \langle c^{\dagger 2} \rangle \langle b^2 \rangle  - \langle{b^\dagger b}\rangle (\langle{c^\dagger c}\rangle + 1) - \langle c^\dagger c \rangle(\langle b^\dagger b \rangle + 1) )
\end{equation}

The average values of all the terms in the R.H.S are henceforth provided below.
\begin{equation}
\langle{b^2}\rangle = \beta^2 e^{-|\beta|^2 (1 - cos4\chi t)} exp[-i2\chi t - i|\beta|^2 sin4\chi t]
\end{equation}  

\begin{equation}
\langle{b^{\dagger 2}}\rangle = \beta^{* 2} e^{-|\beta|^2 (1 - cos4\chi t)} exp[i2\chi t + i|\beta|^2sin4\chi t]
\end{equation}

\begin{equation}
\langle{c^2}\rangle = \gamma^2 e^{-|\gamma|^2 (1 - cos4\chi t)} exp[-i2\chi t - i|\gamma|^2sin4\chi t]
\end{equation}

\begin{equation}
\langle{c^{\dagger 2}}\rangle = \gamma^2 e^{-|\gamma|^2 (1 - cos4\chi t)} exp[i2\chi t + i|\gamma|^2sin4\chi t]
\end{equation}

\begin{equation}
    \langle{b^\dagger b}\rangle = |\beta|^2
\end{equation}

\begin{equation}
    \langle{c^\dagger c}\rangle = |\gamma|^2
\end{equation}

\begin{multline}
    \langle{L_x^2}\rangle = -\frac{1}{4} (\beta^{* 2} e^{-|\beta|^{2} (1 - cos4\chi t)} exp[i2\chi t+ i|\beta|^2 sin4\chi t] \times \gamma^2 e^{-|\gamma|^2(1-cos4\chi t)} exp[-i2\chi t-i|\gamma|^2 sin4\chi t] \\ - \beta^2 e^{-|\beta|^2 (1-cos4\chi t)} exp[-i2\chi t - i|\beta|^2 sin4\chi t] \times \gamma^{* 2} e^{-|\gamma|^2(1-cos4\chi t)} exp[i2\chi t - i|\gamma|^2 sin4\chi t] \\ - |\beta|^2 (|\gamma|^2 + 1) - |\gamma|^2 (|\beta|^2 + 1) )
\end{multline}

\begin{multline}
  4  \langle{L_x^2}\rangle = - \{ e^{-(|\beta|^2 + |\gamma|^2)(1-cos4\chi t)}\} \{ \beta^{* 2} \gamma^2 exp[i(|\beta|^2 - |\gamma|^2) sin4\chi t]\quad \\-\quad \beta^2 \gamma^{* 2} exp[-i(|\beta|^2 - |\gamma|^2)sin4\chi t]  \} + 2|\beta|^2 |\gamma|^2 + |\beta|^2 + |\gamma|^2
\end{multline}

Let's replace $\beta = r_2 e^{i\theta_2}$ and $\gamma = r_3 e^{i\theta_3}$ and rewrite the above expression as:

\begin{multline}
    4 \langle{L_x^2}\rangle = - \{ e^{-(r_2^2 + r_3^2) (1 - cos4\chi t)} \} \{ r_2^2 r_3^2 e^{i2(\theta_3^2 - \theta_2^2)} e^{i(r_2 - r_3) sin4\chi t} \\ - r_2^2 r_3^2 e^{-i2(\theta_3 - \theta_2)} e^{-i(r_2 - r_3) sin4\chi t}  \} + 2 r_2^2 r_3^2 + r_2^2 + r_3^2
\end{multline}

\begin{multline}
    4 \langle{L_x^2}\rangle = - \{ e^{-(r_2^2 + r_3^2) (1 - cos4\chi t)} \} \{ r_2^2 r_3^2 \times 2 \Re{e^{i2(\theta_3^2 - \theta_2^2)} e^{i(r_2 - r_3) sin4\chi t} } \} + 2 r_2^2 r_3^2 + r_2^2 + r_3^2
\end{multline}

And thus we finally arrive at: 
\begin{equation}
4\langle{L_x^2}\rangle = 2 r_2^2 r_3^2 + r_2^2 + r_3^2 - \{ e^{-(1-cos4\chi t)(r_2^2 + r_3^2)} [2r^2_2 r^2_3 cos((r^2_2 - r^2_3 - 2(\theta_2 - \theta_3)sin4\chi t)] \}
\end{equation}
Here, one can clearly see the signatures of the first fractional revival.

\begin{figure}[H]
    \centering
    \includegraphics[width=15cm]{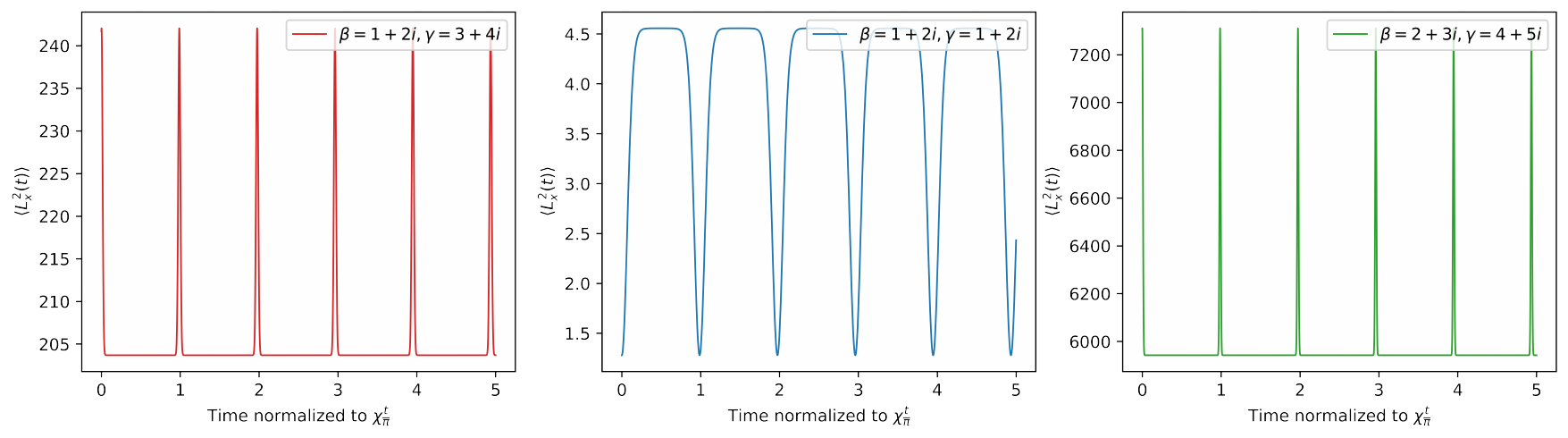}
    \label{fig:true lx2}
    \caption{Plots of $\langle{L_x^2(t)}\rangle$ vs $\frac{\chi t}{\pi}$, for i) $\beta = 1+2j, \gamma = 3+4i$, ii) $\beta = 1 + 2j, \gamma = 1 + 2j$, iii)$\b eta = 2 + 3j, \gamma = 4 + 5j$.}
\end{figure}

\subsubsection{Third and Fourth Moments of $\langle{L_x}\rangle$}

The plots for the higher moments of $L_x$ have been made using the method shown in appendix D. A numerical approach has been undertaken.

The peculiar relation between the values of $\beta$ and $\gamma$ can be seen clearly in the below plot of $\langle{L^3_x}\rangle$. In the second plot, when $\beta = \gamma = 1 + 2i$, the signature of the second fractional revival vanishes completely, and only the full revival can be seen. The manifestations of these fractional revivals observed here, are very much similar to those observed in other observables, excepting the fact that the values taken by the eigen values $\beta$ and $\gamma$ play an important role in the appearance and properties of the signatures of fractional revivals.

\begin{figure}[H]
    \centering
    \includegraphics[width=15cm]{only_lx3.pdf}
    \label{fig:true lx2}
    \caption{Plots of $\langle{L_x^3(t)}\rangle$ vs $\frac{\chi t}{\pi}$, for i) $\beta = 1+2j, \gamma = 3+4i$, ii) $\beta = 1 + 2j, \gamma = 1 + 2j$, iii)$\b eta = 2 + 3j, \gamma = 4 + 5j$.}
\end{figure}

The signatures of the third fractional revival are seen in the plot $\langle{L_x^4}\rangle$
\begin{figure}[H]
    \centering
    \includegraphics[width=15cm]{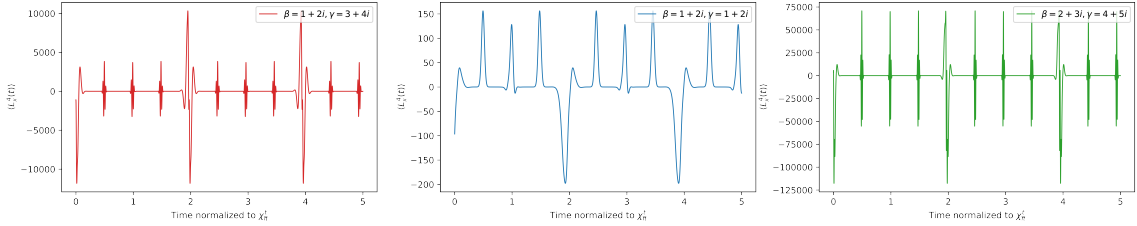}
    \label{fig:true lx2}
    \caption{Plots of $\langle{L_x^4(t)}\rangle$ vs $\frac{\chi t}{\pi}$, for i) $\beta = 1+2j, \gamma = 3+4i$, ii) $\beta = 1 + 2j, \gamma = 1 + 2j$, iii)$\beta = 2 + 3j, \gamma = 4 + 5j$.}
\end{figure}

Even when $\beta = \gamma$ in the second plot for $\langle{L_x^4}\rangle vs time$, the signatures of the third fractional revival do not entirely vanish.

\subsection{Quantum Carpets}

Extending from Talbot Effect, and Talbot carpets, one can construct similar plots, using the state vector $\psi(x, t)$. These so-called "quantum carpets"~\cite{Tooba_frac_rev_carpets} allow us to visualize\cite{Rohith_2015} the probability density function for different instants of time, which gives us a clearer picture of how the system evolves in time. 

\par The basic procedure for the construction of these "carpets" require us to obtain the probability density function $\psi(x)$ for different values of time, and then "placing" them one above the other. The below given "quantum carpet" is constructed, by allowing a coherent state(position representation) of form,
\begin{equation}
    \psi(\alpha = 3, x, t) = e^{-|3|^2} \sum_{n=0}^{50} \frac{3^n}{\sqrt{n!}} \left( \frac{\pi^{\frac{1}{4}}}{\sqrt{2^n n!}}\right) e^{\frac{-x^2}{2}} H_n(x) e^{-in(n-1)\chi t}
\end{equation}
to evolve in time. Here, as is evident, all the constants ($m$, $\omega$, $\hbar$) have been set to unity.

\par Only the first 50 energy eigen-states have been taken into consideration, owing to computational limitations. And hence, the "carpet" appears smudgy and off-focus at certain points. Theoretically, one can avoid such aberrations by taking into consideration a large number of energy eigen states.

\par We know that at the vicinities of time $(n+\frac{j}{k})T_{rev}$ (where $k$ corresponds to the $k^{th}$ fractional revival, and takes values from $1,2,3,..$ and $j=1,2,..,k-1$) the initial wave-packet gets split up into $k$ subpackets which are identical to the initial wavepacket, but with reduced amplitudes. Hence, at these intervals on the plot, one can see smaller manifestations of the initial shape, near $t=0$. The first fractional revival takes place at exactly half the height of the image, where we see two sub-images of the initial image. Similarly at heights one-third and two-thirds of the plot, we see similar but still smaller manifestations of the initial image, which correspond to the second fractional revival where the initial wave-packet gets divided into three sub-packets.  \par Beyond that there is some difficulty in identifying the subsequent fractional revivals, but smaller wavepackets can be seen at times $t = \frac{j}{4}T_{rev}, \frac{l}{5}T_{rev}$, where $j = 1,2,3; l = 1,2,3,4$.    

\begin{figure}[H]
    \centering
    \includegraphics[width=14cm]{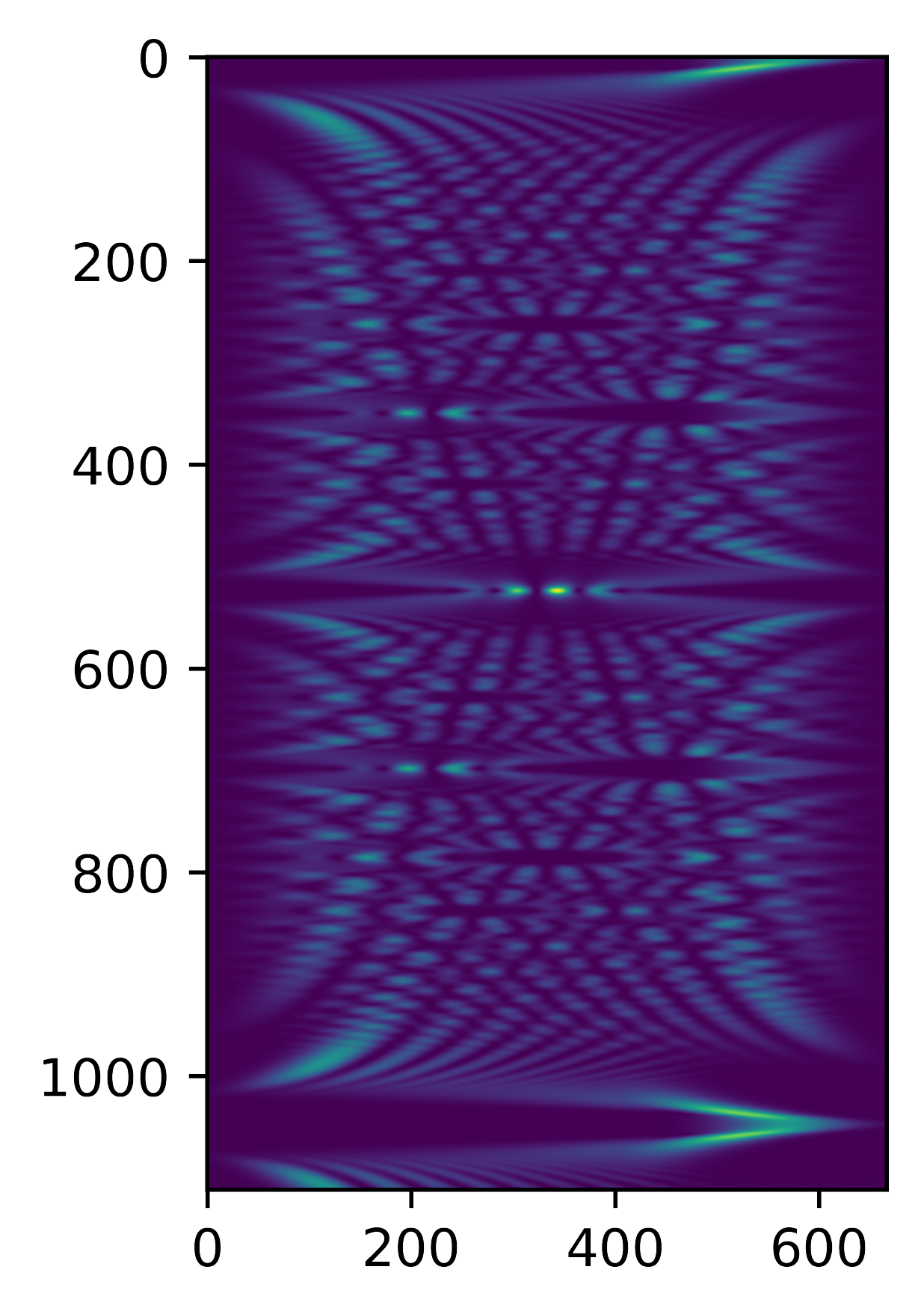}
    \label{fig:true lx2}
    \caption{Quantum Carpet $(|\langle{\psi(x, t)}\rangle|^2)$ constructed for a coherent state in the non-linear kerr Hamiltonian. Here $\alpha = 3$ and $\chi = 10$, and N = 50. All the constants (m, $\omega$, $\hbar$) have been set to unity. The x-axis(position) spans from $-6 \leq x \geq 6$, and the y-axis(time) ranges from $0\leq t \geq 0.\bar{3}$ y-axis. \textit{The values in the axes of the image, correspond to the number of position steps and time-steps used to make this image, and can be promptly ignored.}}
\end{figure}

\begin{figure}[H]
    \centering
    \includegraphics[width=15cm]{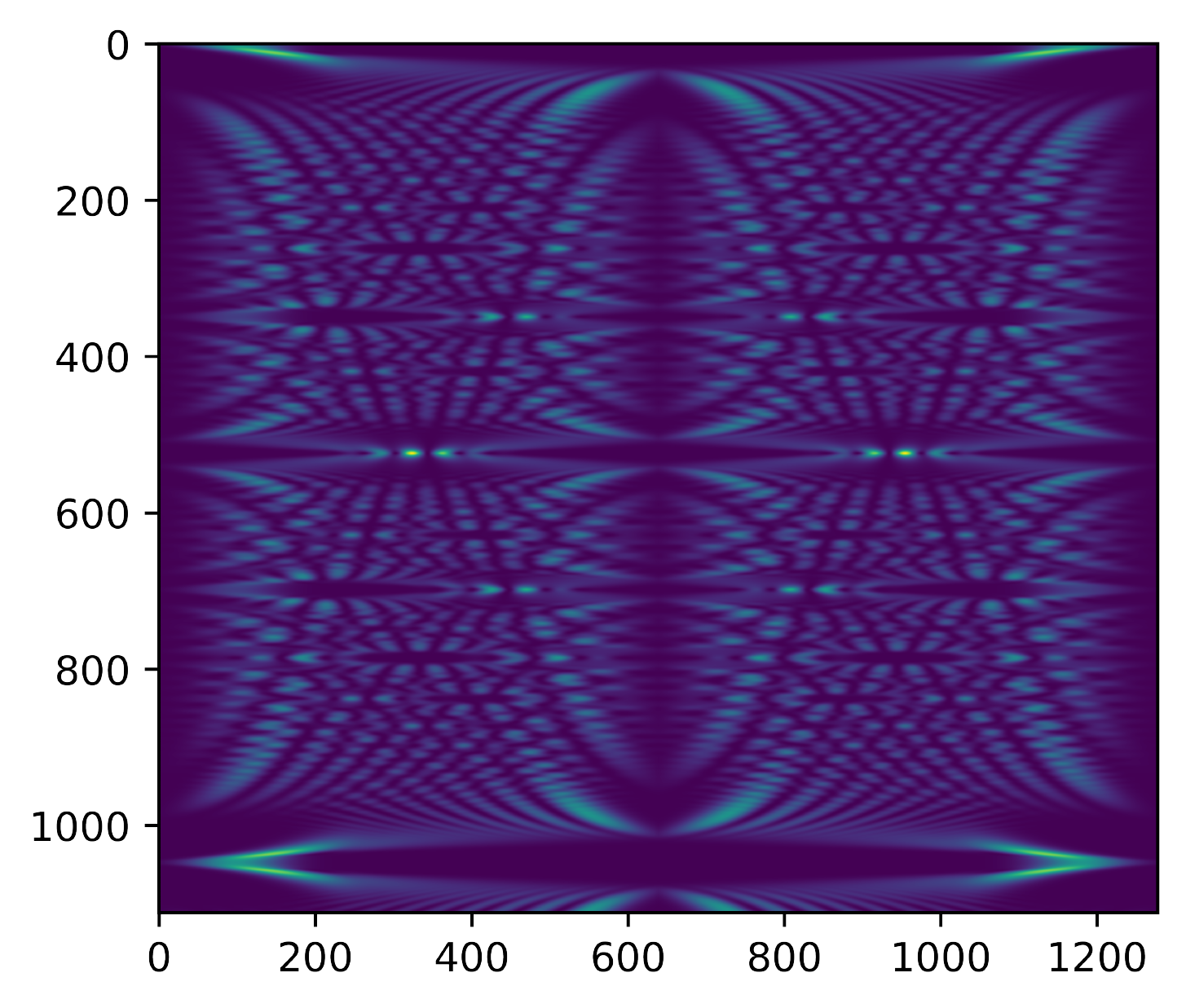}
    \label{fig:true lx2}
    \caption{Similar to the above Quantum Carpet, the pattern has been inverted and added to the left, in order to give a more "complete" image. This has been constructed only for illustrative purposes, and does not correspond to any quantum system.}
\end{figure}

For comparison, below is the "Quantum Carpet" for a coherent state whose time-evolution is governed by an Hamiltonian of form: $\hat{H} = (a^{\dagger}a + \frac{1}{2})\omega$. We see that no dispersion takes place, and as proved by us in section 1.4.4, the coherent state oscillates back and forth, very much similar to a classical particle confined in a harmonic potential. 
 
\begin{figure}[H]
    \centering
    \includegraphics[width=15cm]{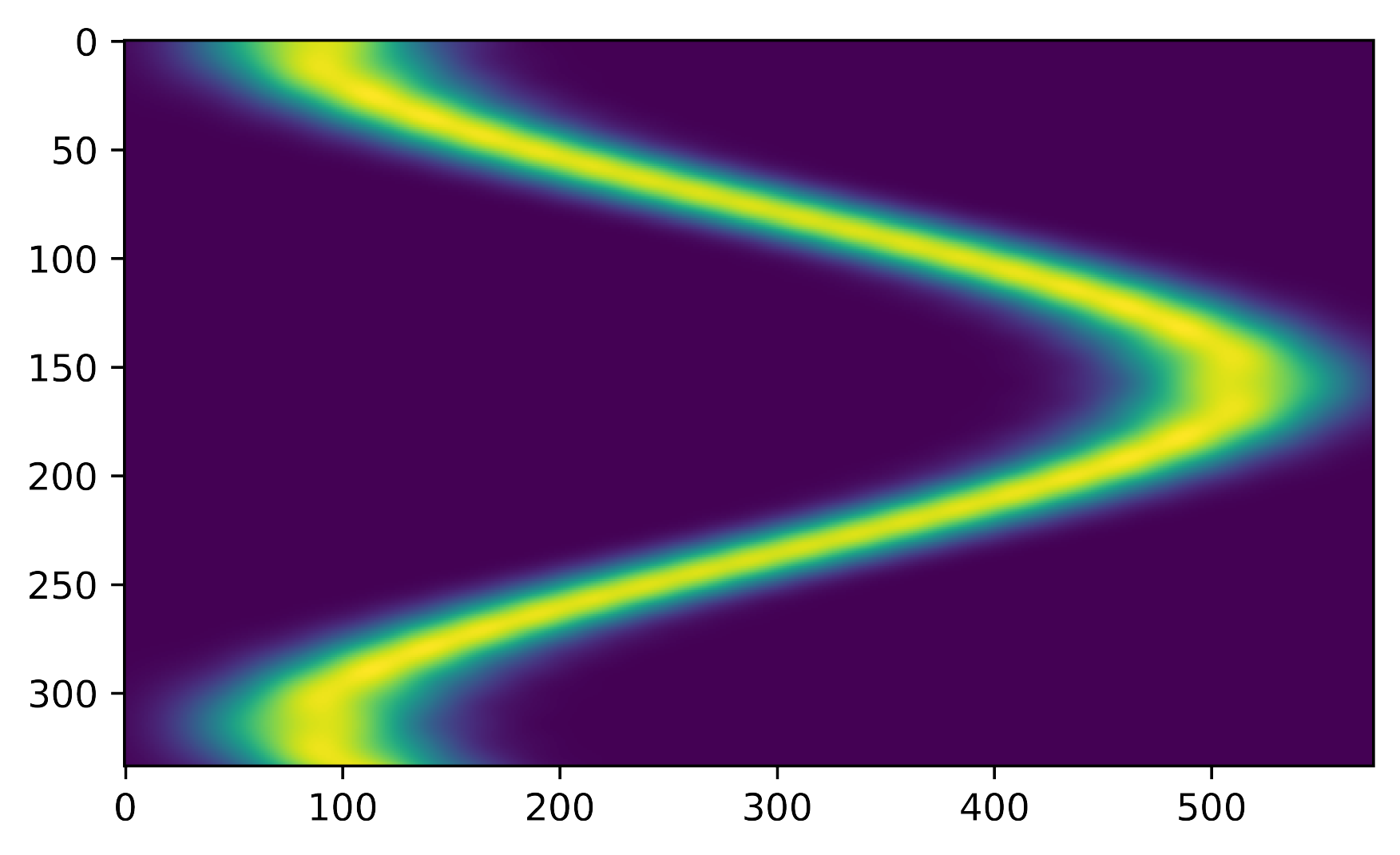}
    \label{fig:true lx2}
    \caption{Quantum Carpet for a coherent state in an Harmonic Oscillator $\hat{H} = \omega(a^{\dagger}a + \frac{1}{2})$}
\end{figure}

\section{Conclusion}
In this work, we have systematically investigated quantum revivals and fractional revivals in systems governed by nonlinear Hamiltonians, with particular emphasis on their manifestation in experimentally accessible observables. Extending earlier studies that focused primarily on position- and momentum-based diagnostics, we have demonstrated that \textbf{angular momentum observables provide a powerful and complementary framework} for identifying fractional revival dynamics. In particular, we have shown that \textbf{higher-order moments of angular momentum components exhibit clear and selective signatures of fractional revivals}, with the order of the moment directly correlating with the order of the fractional revival.

\par Using the Kerr-type nonlinear Hamiltonian as a paradigmatic and analytically tractable model, we derived explicit expressions for the time evolution of angular momentum observables and their moments, and analyzed their behavior alongside standard diagnostics such as the autocorrelation function. Our analysis reveals that while lower-order observables capture full revival dynamics, higher-order angular momentum moments act as sensitive probes of fractional revivals that remain otherwise obscured. Numerical visualizations, including phase-space trajectories and quantum carpets, were employed to provide intuitive insight into the interference
mechanisms underlying revival and fractional revival phenomena.

\par In addition to the quantum analysis, we examined classical analogs of revival phenomena in optical and mechanical systems, and identified structural similarities between quantum fractional revivals and recurrence behavior in classical nonlinear dynamics. Although quantum revivals arise from intrinsically quantum interference effects, these classical analogs offer valuable intuition and underscore the broader universality of revival-like behavior in systems with nonlinear spectra.

\par Taken together, our results broaden the scope of observable-based diagnostics of fractional revivals and establish angular momentum moments as effective, physically transparent indicators of revival dynamics. The framework developed here is applicable to a wide class of quantum systems, including spin systems, quantum optical platforms, and mesoscopic systems with nonlinear evolution. Future directions include the experimental exploration of these signatures and their extension to open quantum systems and higher-dimensional Hilbert spaces.

\section*{Acknowledgements}

The authors would like to sincerely thank Prof. Joseph Prabhakar for his timely help, constant guidance, and insightful discussions throughout the course of this project. His mentorship and encouragement were instrumental in shaping the direction of the work and guiding the author through its various stages.

\section*{Funding}

This research received no specific grant from any funding agency in the public, commercial, or not-for-profit sectors.

\section*{Competing Interests}

The authors declare no competing interests.

% \subsection{FUTURE WORK}

% \begin{enumerate}
%     \item {Investigate spin coherent states, and apply the quantum optimal control algorithms on such states.}\newline
%     \vskip 1em
%     \par The minimum time required for the conversion from one quantum state to another has a lower bound for Hermitian Hamiltonians. The value of this lower bound can be substantially reduced if we consider \textbf{non-hermitian} PT-symmetric Hamiltonians as proposed by Carl.M.Bender et al[25].
%     \item {Such Hamiltonians will be attempted to study and the quantum optimal control algorithms shall be applied to them. }
% \end{enumerate}

\bibliographystyle{unsrt}
\bibliography{references}

\newpage

\appendix
\include{appendix}

\end{document}

%% file: appendix.tex
\section*{APPENDIX}
\section{Analytical form of $a^{\dagger\, r} \, a^{r+s}$}
As the general formula(to obtain the average of $a^{\dagger r}a^{r+s}$) allows us to calculate all the succeeding average values, we find it necessary to provide the exact derivation of the same.\newline
We first venture to find the two following expressions individually:
\begin{equation}
    a^{r + s} \ket{n} = \sqrt{\dfrac{n!}{(n-r-s)!}}       \ket{n-r-s}
\end{equation}

\begin{equation}
    a^{\dagger r} \ket{n - r -s} = \sqrt{\dfrac{(n-s)!}{(n-r-s)!}} \ket{n-s}
\end{equation}
The combined operation of the two above operators on a stationary state $n$ would give:
\begin{equation}
    a^{\dagger r} a^{r + s} \ket{n} = \dfrac{\sqrt{(n-s)!n!}}{(n-r-s)!} \ket{n-s}
\end{equation}
Next, take the coherent state $\ket{\alpha(t)}$:
\begin{equation}
    \ket{\alpha (t)} = e^{\frac{-|\alpha|^2}{2}} \sum ^{\infty}_{n=0} \dfrac{\alpha^n}{\sqrt{n!}} e^{- i \chi n(n-1)t} \ket{n} 
\end{equation}
and operate it with $a^{\dagger r} a^{r+s}$:
\begin{equation}
    a^{\dagger r} a^{r+s} \ket{\psi (t)} = e^{-\frac{|\alpha|^2}{2}} \sum^\infty_{n=0} \frac{\alpha^n \sqrt{(n-s)!}}{(n-r-s)!} e^{-i\chi n(n-1)t} \ket{n-s}
\end{equation}
$\bra{\alpha(t)}$ can be written in the following way:
\begin{equation}
    \bra{\alpha(t)} = e^{-\frac{|\alpha|^2}{2}} \sum ^{\infty}_{m=0} \dfrac{\alpha^{* m}}{\sqrt{m!}} e^{i \chi m(m-1)t} \bra{m} 
\end{equation}
We now calculate the average value of the operator:

\begin{equation}
    \bra{\alpha(t)}a^{\dagger r}a^{r+s}\ket{\alpha(t)} =\langle{a^{\dagger r} a^{r + s}}\rangle = \sum^{\infty}_{m,n = 0} \dfrac{e^{-|\alpha|^2} \sqrt{(n-s)!} \alpha^{n}\alpha^{*m}} {\sqrt{m!}(n-r-s)!}e^{-i\chi n(n-1)t} e^{i\chi m(m-1)t} \langle{m|n-s}\rangle
\end{equation}

where,
\begin{equation}
    \langle{m|n-s}\rangle = \delta_{m,n-s}
\end{equation}
thus,
\begin{equation}
    n-s = m
\end{equation}
\begin{equation}
   \Longrightarrow \quad n = m +s
\end{equation}
   
\begin{equation}
    \langle{a^{\dagger r} a^{r + s}}\rangle = e^{-|\alpha|^2}\sum^{\infty}_{m = 0} \dfrac{ \sqrt{m!} \alpha^{m+s}\alpha^{*m}} {\sqrt{m!}(m-r)!} e^{-i\chi[(m+s)(m+s-1)- m(m-1)] t} 
\end{equation}

As mentioned earlier, let $|\alpha|^2 = \nu$, in order to simplify the succeeding equations.

\begin{equation}
    \langle{a^{\dagger r} a^{r + s}}\rangle = e^{-\nu} \sum^{\infty}_{m = 0} \dfrac{  \alpha^{s}|\alpha|^{2m}} {(m-r)!} e^{-i\chi[(m+s)(m+s-1)- m(m-1)] t} 
\end{equation}

\begin{equation}
    \langle{a^{\dagger r} a^{r + s}}\rangle = e^{-\nu} \alpha^s \sum^{\infty}_{m = 0} \dfrac{|\alpha|^{2m}} {(m-r)!} e^{-i\chi[s^2 + 2ms -s] t} 
\end{equation}

\begin{equation}
    \langle{a^{\dagger r} a^{r + s}}\rangle = e^{-\nu} \alpha^s \sum^{\infty}_{m = 0} \dfrac{|\alpha|^{2m}} {(m-r)!} e^{-i\chi(s(s-1)) t} e^{-2i\chi mst} 
\end{equation}

 The denominator contains a term $(m-r)!$, we see that m must at all instance be either equal to or greater than r, but it can never be lesser than r.(Because that would lead to a negative factorial, which is un-defined) So, we see that $m\geq r$. So, the lower limit is $m=r$.

\begin{equation}
    \langle{a^{\dagger r} a^{r + s}}\rangle = e^{-\nu} \alpha^s \sum^{\infty}_{m = r} \dfrac{|\alpha|^{2m}} {(m-r)!} e^{-i\chi(s(s-1)) t} e^{-2i\chi mst} 
\end{equation}

Let us define $m-r=h$.

Thus h goes from 0 to $\infty$.

\begin{equation}
    \langle{a^{\dagger r} a^{r + s}}\rangle = e^{-\nu} \alpha^s e^{-i\chi(s(s-1)) t} \sum^{\infty}_{h=0} \dfrac{\nu^{h+r}} {h!}  e^{-2i\chi (h+r)st} 
\end{equation}

\begin{equation}
    \langle{a^{\dagger r} a^{r + s}}\rangle = e^{-\nu} \alpha^s e^{-i\chi(s(s-1)) t} \sum^{\infty}_{h=0} \dfrac{\nu^{h}} {h!}  {e^{(-2i\chi st)}}^{h}  {e^{(-2i\chi st)}}^{r} \nu^r
\end{equation}
The running index here is $h$, and so one can remove all the terms containing $r$ outside the summation and write the remaining terms in the form of $\sum^{\infty}_{n=0} \frac{\text{some value}^n}{n!}$ 
\begin{equation}
    \langle{a^{\dagger r} a^{r + s}}\rangle = e^{-\nu} \alpha^s \nu^r e^{-i\chi(s(s-1)) t}  {e^{(-2i\chi st)}}^{r}  \sum^{\infty}_{h=0} \dfrac{{(\nu e^{-2i\chi st})}^{h}} {h!}
\end{equation}
which we next convert into an exponential function.

\begin{equation}
    \langle{a^{\dagger r} a^{r + s}}\rangle = e^{-\nu} \alpha^s \nu^r e^{-i\chi(s(s-1) +2rs) t}  exp(\nu e^{-2i\chi st})
\end{equation}
Expanding the final term, using Euler's Formula, we can write it as:
\begin{equation}
    \langle{a^{\dagger r} a^{r + s}}\rangle = e^{-\nu} \alpha^s \nu^r e^{-i\chi(s(s-1) +2rs) t}  e^{\nu cos2\chi st} e^{-i\nu sin2\chi st}
\end{equation}

\begin{equation}
    \Longrightarrow \langle{a^{\dagger r} a^{r + s}}\rangle = \alpha^s \nu^r e^{-\nu} e^{\nu cos2\chi st} e^{-i\chi(s(s-1) +2rs) t}  e^{-i\nu sin2\chi st}
\end{equation}
And so, finally,
\begin{equation}
    \langle{a^{\dagger r} a^{r + s}}\rangle = \alpha^s \nu^r e^{-\nu(1- cos2\chi st)} e^{[-i\chi(s(s-1) +2rs)t-i\nu sin2\chi st]}
\end{equation}

\section{Auto-correlation function Calculation}

The following program was used to plot the absolute square of the auto-correlation function for a coherent state, whose time evolution is governed by the non-linear Kerr Hamiltonian.

At first, we'd like to create any sub functions which we will use for the final plot creation. From section 1.6, we can write the auto-correlation function as:

\begin{equation}
    \bra{\alpha(t)}\ket{\alpha(0)} = e^{-|\alpha|^2} \sum_{n=0}^{\infty} \frac{|\alpha|^{2n}}{{n!}} e^{-i(n(n-1))\chi t} 
\end{equation}

The above inner product will clearly give us a complex number as an output and so, usually its absolute function is plotted. (Although one can always plot the real part and the imaginary part separately.)

Let's first define a function which returns the coefficients of each energy eigen state in the coherent state, when we specify the average photon amplitude $\alpha$. We'd like to re-emphasize that the coherent state is:

\begin{equation}
    \ket{\alpha(t)} = e^{-\frac{|\alpha|^2}{2}} \sum^{\infty}_{n=0} \frac{\alpha^n}{\sqrt{n!}} e^{-i\frac{\hat{H}t}{\hbar}} \ket{n} 
\end{equation}

One thing clear from the expression of a coherent state is that, for large $n$, the coefficients can be neglected owing to the presence of $\sqrt{n!}$ in the denominator. And as we are approaching the problem analytically, we'd like to promptly take into consideration only those terms which contribute to the end result in a non-negligible manner.

Hence, as of now, we shall confine ourselves to a finite number of stationary states.

That is, if I wish to work with a coherent state whose average photon amplitude is $\alpha$, and the coherent state is a superposition of N stationary states, then upon specifying as such, the function must return to us an array of the coefficients $c_n$.

We know that $\hat{H}=\hat{H}(n)$, and for a non-linear Kerr Hamiltonian of form $a^{\dagger}a \hbar \chi$, $\hat{H} \propto n(n-1)$ and for a linear harmonic oscillator , $\hat{H}\quad \alpha \quad n$. This determines the time evolution of the coherent state, and in our program is imbibed within the variable $\verb|time_param|$.

 Whether the coherent state is annotated as a bra vector or a ket vector needs to be specified, as the final parameter next to the instant of time $\verb|t|$ at which we wish to obtain the coefficients.

\begin{tcolorbox}   
\begin{lstlisting}[language = Python]

import numpy as np
import matplotlib.pyplot as plt
import cmath
from math import *

chi = 10 / np.pi 
#frequency term, divided by pi for normalization.

def coherent(N, p, q, t, state):
    alpha = complex(p, q)
    mod_alpha = cmath.polar(alpha)[0]
    co1 = np.exp(-0.5 * mod_alpha**2)
    coherent_array = []
    for n in range(N + 1):
        if state == "bra":
            co2 = (alpha.conjugate())**n / np.sqrt(factorial(n))
        elif state == "ket":
            co2 = alpha**n / np.sqrt(factorial(n))
        time_param = n * (n - 1)
        t1 = np.cos(time_param * chi * t)
        t2 = np.sin(time_param * chi * t)
        time = complex(t1, t2)
        fin_val = co1 * co2 * time
        coherent_array.append(fin_val)

    return np.array(coherent_array)

\end{lstlisting}
\end{tcolorbox}

 Our expected output shall be an array of values which look like:
\begin{equation}
    \left(\begin{array}{*{11}c}c_0 & c_1 & c_2 & c_3 & c_4 & c_5 & c_6 & c_7 & c_8 & c_9 & ... \end{array}\right)
\end{equation}
where $c_n$ is calculated using the following formula:
\begin{equation}
    c_n = e^{-\frac{|\alpha|^2}{2}} \frac{\alpha^n}{\sqrt{n!}}
\end{equation}
Let us check if the above snippet provides us with the correct coefficients $c_n$. We shall set $\alpha=p+ij=1+1j$ and $\verb|N=20|$ and time $\verb|t=0|$.

\begin{tcolorbox}
\begin{lstlisting}[language = Python]
initial_state = coherent(20, 1, 1, 0, "ket")
print(initial_state)
\end{lstlisting}
\end{tcolorbox}

\begin{tcolorbox}
\begin{lstlisting}[language = Python]

#OUTPUT

array([ 3.67879441e-01+0.00000000e+00j,  3.67879441e-01+3.67879441e-01j,
        0.00000000e+00+5.20260095e-01j, -3.00372306e-01+3.00372306e-01j,
       -3.00372306e-01+0.00000000e+00j, -1.34330579e-01-1.34330579e-01j,
        0.00000000e+00-1.09680458e-01j,  4.14553167e-02-4.14553167e-02j,
        2.93133355e-02+0.00000000e+00j,  9.77111184e-03+9.77111184e-03j,
        0.00000000e+00+6.17979374e-03j, -1.86327792e-03+1.86327792e-03j,
       -1.07576401e-03+0.00000000e+00j, -2.98363253e-04-2.98363253e-04j,
        0.00000000e+00-1.59481867e-04j,  4.11780410e-05-4.11780410e-05j,
        2.05890205e-05+0.00000000e+00j,  4.99357096e-06+4.99357096e-06j,
        0.00000000e+00+2.35399193e-06j, -5.40042785e-07+5.40042785e-07j])

\end{lstlisting}
\end{tcolorbox}

Hence the program works fine.

Finally, we write a function \verb|correlation(real part, imaginary part)| to compute the auto-correlation function.

% \begin{tcolorbox}
\begin{lstlisting}[language = Python]
steps = np.linspace(0, 1, 10000)

def correlation(p ,q):
    modacf = []
    for t in np.arange(0, 1, 0.0001):
        curr_state = co(20, p, q, t, "bra")
        ac_val = 0
        for n in range(len(ini_state)):
            ac_val += curr_state[n] * initial_state[n]
        modacf.append((cmath.polar(ac_val)[0])**2)  
    return modacf
    
final_array = correlation(1, 1)
plt.plot(steps, final_array)
plt.show()
\end{lstlisting}
% \end{tcolorbox}

Now, all that remains is to create an array for the x-axis(time), and the output array of the function can be stored, which we can then use for the plot.
One can tweak the value of the variable $\verb|time_param|$ in the above program to be different functions of $n$, and obtain plots for several other stationary state dependent hamiltonians.

\section{Links to Animations}

This section of the appendix shall provide all the links for any animations which are discussed in the dissertation.

\subsection{Dancing pendulums}

\begin{enumerate}
    \item A set of 15 uncoupled pendulums with monotonically increasing lengths.
\begin{verbatim}
https://www.glowscript.org/#/user/p.b.ashish786/folder/Project
/program/dancingpendulums
\end{verbatim}
    
    \item A set of 50 uncoupled pendulums with monotonically increasing lengths.(More suitable if the reader wishes to see prominent patterns of fractional revivals).
\begin{verbatim}
https://www.glowscript.org/#/user/p.b.ashish786/folder/Project
/program/dancingpendulums2
\end{verbatim}
\end{enumerate}

\subsection{"Phase plots of $\langle{x}\rangle$ vs $\langle{p}\rangle$}
The below provided links can be accessed to view the animated phase plots of $\langle{x(t)}\rangle$ vs $\langle{p(t)}\rangle$.

Environment Used: \verb|VPython GlowScript IDE|

\begin{enumerate}

\item For $p_1 = 1$ and $q_1=1$:

\begin{verbatim}
https://www.glowscript.org/#/user/p.b.ashish786/folder/Project
/program/%3Cx%3Evs%3Cp%3Emu1
\end{verbatim}

\item For $p_1 = \sqrt{10}$ and $q_1 = \sqrt{10}$:

\begin{verbatim}
https://www.glowscript.org/#/user/p.b.ashish786/folder/Project
/program/%3Cx%3Evs%3Cp%3Emusqrt10
\end{verbatim}

\item For $p_1=10$ and $q_1=10$:

\begin{verbatim}
https://www.glowscript.org/#/user/p.b.ashish786/folder/Project
/program/%3Cx%3Evs%3Cp%3Emu10
\end{verbatim}
\end{enumerate}

\section{Angular Momentum Observable - Higher Moments computation using Python}
\quad Owing to the complexity involved in the calculation of the average values of $L_x^3$ and $L_x^4$, Python language is used to write a program to plot the average of the respective values vs. time. Here, only the function (\verb|aadag| , short for \verb|a, a dagger|) employed to calculate the general formula for $a^{\dagger k}a^{k+l}$ is given. It is assumed that the reader is familiar with the \verb|matplotlib| module, which is used in conjuction with the below given code snippet to obtain the plots.\newline  By changing the input arguments given to the function and finding the complex conjugates of the output values(when necessary), one can thus obtain the required values for calculation.\newline Modules used: \verb|numpy, cmath|.\newline Function: \verb|aadag(p = real part, q = imaginary part, {k , l} = power values,|
\verb|t = time)|
\newline

\begin{tcolorbox}
\begin{lstlisting}[language = Python]
import numpy as np
import cmath

f = 5 #chi value (in our case, some positive constant)

def aadag(p,q,k,l,t):               
    
    z = complex(p,q) 
    #gives us a complex number z = p + iq
    w = (z.polar[0])**2 / 2
    #z.polar returns a tuple of form z.polar[0] = modulus
    #of z,z.polar[1] = angle
    exppow = np.exp(-1 * w * (1 - np.cos(2 * l * f * t)))
    inside = (l * (l - 1)) + (2 * k * l)
    inside2 = w * np.sin(2 * l * f * t)
    inside3 = 2 * f * inside * t + inside2
    compval = complex(np.cos(inside3), np.sin(inside3))
    finalvalue = (z**l) * (w**k) * exppow * compval
    return finalvalue

\end{lstlisting}
\end{tcolorbox}

In this simple snippet written above, for ease of calculation and to reduce the abstraction, certain variables have been assigned to certain parts of the general formula to obtain the expression $\langle{a^{\dagger r}a^{r+s}}\rangle$

For example, let us take $\langle{L_x^3}\rangle$, which can be written in terms of the operators $b$,$b^\dagger$,$c$,$c^\dagger$ as:
\begin{equation}
L_x^3 = \frac{-1}{8i} (b^\dagger c - c^\dagger b )^3
\end{equation}
Upon expansion, we get:
\begin{multline}
    L_x^3 = \frac{-1}{8i} (b^{\dagger 3}c^3 - b^{\dagger 2}b c^2c^\dagger - b^\dagger bb^\dagger cc^\dagger c + b^\dagger b^2 c c^{\dagger 2} \\ - c^\dagger c^2 b b^{\dagger 2} - c^\dagger c c^\dagger bb ^\dagger b + c^{\dagger 2}c b^2b^\dagger - c^{\dagger 3}b^3) 
\end{multline}
\par
Operators are position-specific and the order of their multiplication is important, in the case of raising and lowering operators corresponding to the same axis or basis($b$,$b^\dagger$ and $c$, $c^\dagger$).
Moreover, the general formula at our disposal, only gives us the average values for combinations of the raising and lowering operators in the form of $a^{\dagger x} a^y$ and not for combinations in the form of $a^x a^{\dagger y}$.\newline \newline \par Hence, we employ the commutation relation $[a, a^\dagger] = 1$ to modify the expressions in such a manner that they can be fit into the function.

\begin{multline}
    L_x^3 = \frac{-1}{8i} (b^{\dagger 3}c^3 - (b^{\dagger 2}b c^\dagger c^2 + 2c) - (b^{\dagger 2}b + b^\dagger)(c^\dagger c^2 + c) + b^\dagger b^2 (c^{\dagger 2}c + 2c^\dagger) - c^\dagger c^2(b^{\dagger 2}b + 2b^\dagger)\\ - (c^{\dagger 2}c + c^\dagger)(b^\dagger b^2 + b^\dagger)
    +c^{\dagger 2}c(b^\dagger b^2 + 2c) - c^{\dagger 3}b^3)
\end{multline}

After computing the above equation, one only needs to alter the necessary values of k, l to obtain the required values. \newline For instance, to obtain the value of say, the term $b^{\dagger}b^2$ (where $k = 1$,$l =1$) at some time $t = 0.5$ and for $\alpha = \sqrt{10} + \sqrt{10}i$, we can write the following:
\newline

\begin{tcolorbox}
\begin{lstlisting}[language = Python]
some_var = aadag(np.sqrt(10), np.sqrt(10),1 ,1 ,0.5)
print(some_var)
\end{lstlisting}
\end{tcolorbox}
The above snippet of code, though amateurish in nature,helps us to numerically obtain the expression $\langle{a^{\dagger k} a^{k+l}(t)}\rangle$ allowing us to compute the average values of any operator which can be written using the raising and lowering operators.